\documentclass[aps,twocolumn,nofootinbib,groupedaddress,superscriptaddress,longbibliography,notitlepage]{revtex4-2}

\usepackage{amsmath,amssymb}
\usepackage[normalem]{ulem}
\usepackage{graphicx}
\usepackage{longtable,multirow}
\usepackage{mathtools}
\usepackage{dsfont}
\usepackage{amsfonts}
\usepackage[dvipsnames]{xcolor}
\usepackage{url}
\usepackage{bbold}

\usepackage[colorlinks=true,linkcolor=RoyalBlue,citecolor=RoyalBlue,urlcolor=RoyalBlue,hypertexnames=false]{hyperref}
\newcommand{\nocontentsline}[3]{}
\newcommand{\tocless}[2]{\bgroup\let\addcontentsline=\nocontentsline#1{#2}\egroup}

\def\ba#1\ea{\begin{align}#1\end{align}}
\def\bg#1\eg{\begin{gather}#1\end{gather}}
\def\bpm{\begin{pmatrix}}
\def\epm{\end{pmatrix}}
\def\bbm{\begin{bmatrix}}
\def\ebm{\end{bmatrix}}

\newcommand{\nn}{\nonumber \\ }
\newcommand{\bb}[1]{{\mathbf #1}}
\newcommand{\bs}[1]{{\boldsymbol #1}}
\newcommand{\bx}{\bb x}
\newcommand{\bk}{\bb k}

\newcommand{\bq}{\bb q}

\newcommand{\bR}{\bb R}

\newcommand{\bG}{\bb G}

\newcommand{\td}[1]{\widetilde{#1}}
\newcommand{\cm}{\overline}
\newcommand{\mc}[1]{\mathcal{#1}}

\newcommand{\der}{\partial}
\newcommand{\dg}{\dagger}

\newcommand{\sg}{\sigma}

\newcommand{\ep}{\epsilon}
\newcommand{\Z}{\mathbb{Z}}

\newcommand{\C}{\mathbb{C}}

\newcommand{\ket}[1]{|#1\rangle}
\newcommand{\bra}[1]{\langle#1|}
\newcommand{\brk}[2]{\langle#1|#2\rangle}
\newcommand{\kbr}[2]{|#1\rangle\langle#2|}

\newcommand{\Det}{{\rm Det}}
\newcommand{\Ncell}{{N_{\rm cell}}}
\newcommand{\Norb}{{N_{\rm orb}}}

\newcommand{\themecol}[1]{{\color{RoyalBlue} #1}}

\newcommand{\ourtitle}{Chern Numbers from Symmetry-Enforced Zeros and Dispersion Shifts}

\allowdisplaybreaks

\begin{document}
\title{\ourtitle}

\author{Yoonseok Hwang}
\author{Henry Davenport}
\author{Frank Schindler}
\affiliation{Blackett Laboratory, Imperial College London, London SW7 2AZ, United Kingdom}

\begin{abstract}
Wave-function zeros and their local vorticities provide a complementary characterization of Chern topology to more commonly used Berry-curvature formulations.
Using wave-function overlaps between a Chern band and freely chosen reference states, we show that crystalline symmetry can enforce their zeros and fix their vorticities modulo the rotation order from the angular momenta of the band and reference state.
We further show that weak local perturbations produce band-dispersion shifts proportional to the absolute square of the overlap, allowing its zeros and local structure to be read off directly from the dispersion.
Combined with crystalline symmetry, this can identify Chern topology without measuring wave-function phases.
Our construction applies to multicomponent and multiband wave functions and allows reference states whose momentum-space representation may vanish at some momenta.
This permits compact localized states to be used directly as reference states and makes use of their known real- and momentum-space properties in flat-band systems.
The dispersion shift near an overlap zero further provides a lower bound on the quantum metric at the zero.
\end{abstract}

\maketitle

\let\oldaddcontentsline\addcontentsline
\renewcommand{\addcontentsline}[3]{}

Chern bands describe the integer quantum Hall effect and can also arise in lattice systems without a net magnetic field~\cite{thouless1982quantized,haldane1988model,hasan2010colloquium,xiao2010berry}.
More recently, they have attracted intense interest with the discovery of integer and fractional quantum anomalous Hall effects in two-dimensional quantum materials~\cite{bistritzer2011moire,cao2018unconventional,cao2018correlated,sharpe2019emergent,serlin2020intrinsic,regnault2011fractional,neupert2011fractional,sheng2011fractional,zhang2019nearly,cai2023signatures,zeng2023thermodynamic,park2023observation}.
Conventionally, their topology is characterized by the Berry curvature, whose Brillouin-zone integral gives the Chern number~\cite{xiao2010berry}.
A complementary geometric characterization relates the first Chern class of a line bundle to the zeros of its sections~\cite{griffiths2014principles,faure2000topological}.
In quantum Hall systems, wave-function zeros and their phase vorticities were already related to the Chern number in early studies~\cite{kohmoto1985topological,hatsugai1993chern}.
Wave-function zeros and singularities in momentum space have since appeared in flat-band systems~\cite{rhim2020quantum,bergman2008band,rhim2019classification,hwang2021general,graf2021designing,hwang2021wave,hwang2021flat}, symmetry-enforced zeros of Bloch-state components~\cite{hwang2021flat,holbaek2026protection} and of exciton wave functions~\cite{hwang2026stable,hwang2026exciton,paiva2024shift}.
Related ideas have also appeared for zeros involving real-space coordinates~\cite{silva2021topological,ingham2026real} and in approaches based on vortexability and generalized Landau levels~\cite{ledwith2023vortexability,fujimoto2025higher,liu2025theory}.

For a general multicomponent Bloch state $\ket{u (\bk)}$, however, there is no unique scalar amplitude whose zeros can be defined.
One can instead introduce a reference state $\ket{v (\bk)}$ and consider its scalar overlap with $\ket{u (\bk)}$.
Ref.~\cite{li2024constraints} used a trial Wannier state as such a reference to relate the total overlap vorticity to the Chern number and to characterize the Wannier obstruction and real-space representations of Chern bands.
Several questions remain for crystalline systems.
The role of crystalline symmetry in enforcing overlap zeros and constraining their vorticities remains unclear.
More general reference states may themselves carry nontrivial topology, or may even vanish at some momenta, as can occur for Fourier transforms of compact localized states (CLSs)~\cite{bergman2008band,rhim2019classification,hwang2021flat}.
It is also desirable to access the vorticities through observables without measuring wave-function phases.

In this work, we study momentum-space overlap zeros between band eigenstates and reference states and their associated vorticities.
We show that crystalline symmetry can enforce such zeros at high-symmetry momenta and constrain their vorticities through symmetry eigenvalues.
The framework applies to both normalized and unnormalized reference states.
The latter are taken to be smooth and globally defined over the first Brillouin zone (BZ), but may vanish at some momenta, allowing CLSs to be used directly.
For such localized reference states, local perturbations constructed from the reference produce dispersion shifts proportional to the absolute square of the overlap.
Their behavior near a zero, combined with crystalline symmetry, can determine the vorticity without measuring wave-function phases and, in favorable cases, the Chern number.

\themecol{\it Overlap zeros and vorticity---}
Before considering crystalline symmetry, we first formulate wave-function zeros for multicomponent states through their overlaps with reference states.
All states are represented in the same orbital basis and obey the corresponding boundary identification over the BZ, as specified below.
Throughout, any normalized state is assumed to define a smooth projector over the BZ, so that its Chern number is well defined.

We first consider an isolated Chern band with a normalized eigenstate $\ket{u (\bk)}$ of the Bloch Hamiltonian, satisfying $\brk{u (\bk)}{u (\bk)}=1$ at every $\bk$.
We also introduce a reference state $\ket{w (\bk)}$, which need not be normalized and may vanish at some momenta.
The multiband generalization is discussed in the Discussion.
We take $\ket{w (\bk)}$ to be smooth over the BZ.
We first focus on $\ket{w (\bk)}$ and return below to the case of a normalized reference state.

We define the overlap $\zeta_w (\bk) = \brk{w (\bk)}{u (\bk)}$.
The overlap construction with a trial Wannier reference was considered in Ref.~\cite{li2024constraints}.
Here we allow the reference state itself to vanish at some momenta, and treat such zeros on the same footing as other overlap zeros when analyzing their vorticities and crystalline constraints.
In two dimensions (2D), the conditions ${\rm Re} [\zeta_w] = {\rm Im} [\zeta_w] = 0$ generically define isolated point zeros, as remains the case for the crystalline symmetries considered below.
Extended zero sets can be treated in the same way by integrating around their boundary, as discussed in the Supplemental Material (SM)~\cite{supple}.
Around each isolated zero $\bk_s$, we choose a small disk $D_s$ and a smooth local gauge for $\ket{u (\bk)}$, and define
\bg
\nu (\bk_s) = \frac{1}{2 \pi} \oint_{\der D_s} d \arg \zeta_w (\bk).
\label{eq:vort_def}
\eg
Here and below, $d f(\bk) = d \bk \cdot \der_\bk f(\bk)$.

We next rewrite Eq.~\eqref{eq:vort_def} in a gauge-invariant expression.
Let $\mc A_u (\bk) = i \brk{u (\bk)}{d u (\bk)}$ be the Berry connection of $\ket{u (\bk)}$, and define a gauge-invariant one-form
\bg
\Omega_w (\bk) = d \arg \zeta_w (\bk) + \mc A_u (\bk),
\label{eq:om_def}
\eg
away from the zeros of $\zeta_w(\bk)$.
Under a gauge transformation of $\ket{u (\bk)}$, with $\ket{w (\bk)}$ kept fixed, the two terms transform oppositely, so $\Omega_w$ is gauge invariant.
Moreover, $d \Omega_w = \mc F_u$, which is the Berry-curvature 2-form.
Note that the term $d \arg \zeta_w$ can also be interpreted as a singular connection, closely related to that of Refs.~\cite{mera2022singular,paiva2024shift}, as discussed in the SM~\cite{supple}.

Since $\ket{u (\bk)}$ has a smooth projector, a smooth gauge can be chosen on $D_s$, so $\oint_{\der D_s} \mc A_u (\bk) = \int_{D_s} \mc F_u = O(\ep^2)$ vanishes as the radius $\ep$ of $D_s$ goes to zero.
Equations~\eqref{eq:vort_def} and \eqref{eq:om_def} then give
\bg
2 \pi \nu (\bk_s) = \lim_{\ep \to 0} \oint_{\der D_s} \Omega_w (\bk),
\label{eq:vort_def2}
\eg
which can be evaluated numerically using the discretized formula given in the SM~\cite{supple}.

\themecol{\it Vorticity sum rule and Chern number---}
The total vorticity of the overlap zeros obeys
\bg
\nu_{\rm tot} [\zeta_w] = \sum_s \, \nu (\bk_s) = - \mc C_u,
\label{eq:sum_w}
\eg
where $\mc C_u$ is the Chern number of $\ket{u (\bk)}$.
No Chern number associated with $\ket{w (\bk)}$ appears.
If $\ket{w (\bk)}$ is nonzero everywhere, its normalized state has zero Chern number.
If it vanishes, such a Chern number need not be defined.
Nevertheless, the sum includes zeros inherited from zeros of the reference state $\ket{w (\bk)}$.
Such zeros need not have zero vorticity and can therefore contribute nontrivially to the sum rule, as illustrated with an example in the SM~\cite{supple}.
To see Eq.~\eqref{eq:sum_w}, let $M$ denote the BZ with small disks around all zeros of $\zeta_w$ removed.
Since $\Omega_w$ is gauge invariant and $d \Omega_w = \mc F_u$, Stokes' theorem and Eq.~\eqref{eq:vort_def2} give $2 \pi \nu_{\rm tot} [\zeta_w] = - \lim_{\ep \to 0} \oint_{\der M} \Omega_w (\bk) = - \lim_{\ep \to 0} \int_M \mc F_u$.
Thus, as the disks shrink, $\nu_{\rm tot} [\zeta_w] = - \frac{1}{2\pi} \int_{\rm BZ} \mc F_u (\bk) = - \mc C_u$.

For a normalized reference state $\ket{v (\bk)}$ with possibly nonzero Chern number $\mc C_v$, the corresponding gauge-invariant one-form is $\Omega_v (\bk) = d \, {\rm arg} \, \zeta_v (\bk) + \mc A_u (\bk) - A_v (\bk)$, where $\zeta_v (\bk) = \brk{v (\bk)}{u (\bk)}$.
The same argument gives~\cite{supple}
\bg
\nu_{\rm tot} [\zeta_v] = \mc C_v - \mc C_u,
\label{eq:sum_v}
\eg
The case of a topologically trivial trial Wannier reference ($C_v=0$) was considered in Ref.~\cite{li2024constraints}.
Allowing both normalized and unnormalized references is useful when crystalline symmetry restricts the available choices.
A symmetry-compatible reference may carry nonzero Chern number or vanish at some momenta~\cite{supple}.

\themecol{\it Symmetry-enforced zeros and vorticities---}
We now consider crystalline symmetry.
Since its action depends on the basis orbitals and their positions, we first specify our convention.
Let $\ket{\bR, \alpha}$ denote a basis orbital at $\bR + \bx_\alpha$, with $\bR$ a Bravais lattice vector, and define $V(\bk) = {\rm Diag} (e^{i \bk \cdot \bx_1}, \cdots, e^{i \bk \cdot \bx_\Norb})$.
Then, the Hamiltonian satisfies $H(\bk+\bG) = V(\bG)^{-1} H(\bk) V(\bG)$, where $\bG$ is a reciprocal-lattice vector.
We use the periodic gauge~\cite{alexandradinata2014wilson,shiozaki2017topological}, in which the states satisfy $\ket{a (\bk + \bG)} = V(\bG)^{-1} \ket{a (\bk)}$ for $a=u,v,w$.
This boundary identification makes $\zeta_{w,v} (\bk)$, $\Omega_{w,v} (\bk)$, and the sewing phases $\xi_{a,g} (\bk)$ defined below periodic in the BZ.
A particularly useful choice for $\ket{w (\bk)}$ is the Fourier transform of a CLS, which provides a compactly localized reference in real space.
For such a CLS reference, its components $\cm{w}_\alpha (\bk) = e^{i \bk \cdot \bx_\alpha} w_\alpha (\bk)$ are Laurent polynomials in $e^{i \bk \cdot \bb a_i}$ with $\bb a_i$ the primitive lattice vectors~\cite{chen2014impossibility,dubail2015tensor,read2017compactly,read2017compactly,rhim2019classification,hwang2021flat,hwang2021general}.
More details on these conditions are summarized in the SM~\cite{supple}.
%

\begin{figure*}[t]
\centering
\includegraphics[width=0.99\textwidth]{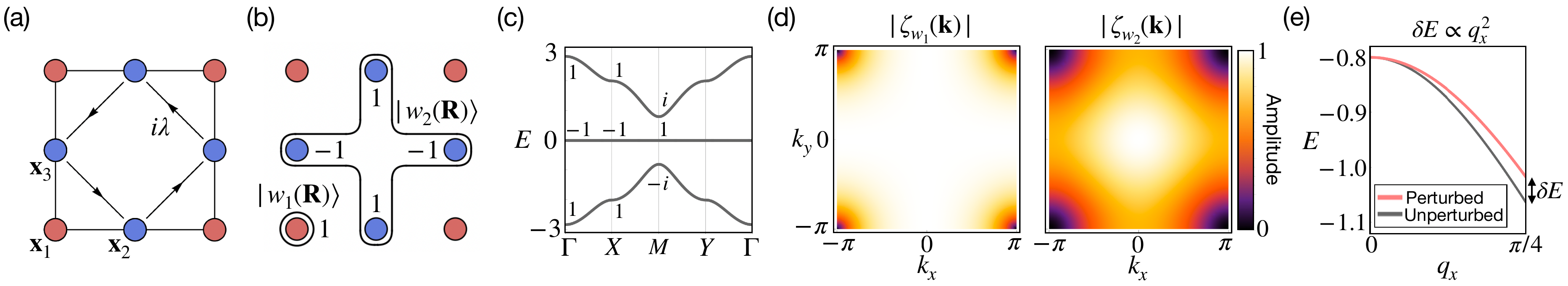}
\caption{Overlap zeros, vorticities, and dispersion shifts in the Lieb lattice model.
(a) Unit cell, with solid lines denoting nearest-neighbor hopping and arrows denoting spin-orbit coupling.
(b) CLSs $\ket{w_{1,2} (\bR)}$ corresponding to the reference states $\ket{w_{1,2} (\bk)}$, with their amplitudes indicated.
(c) Band structure and symmetry eigenvalues.
(d) $|\zeta_{w_{1,2}} (\bk)|$, each normalized to unit maximum.
Both vanish at $M$.
The zero of $\zeta_{w_1} (\bk)$ is symmetry enforced, with $\nu (M) = 1$ and linear behavior, whereas that of $\zeta_{w_2} (\bk)$ originates from $\ket{w_2 (\bk)}$ itself, with $\nu (M)=0$ and quadratic behavior.
(e) Dispersion near $M$ in the $M \to X$ direction, $\bk = (\pi-q_x,\pi)$, before (black) and after (red) adding $\eta \kbr{w_1 (\bk)}{w_1 (\bk)}$ with $\eta = 0.2$.
The quadratic dispersion shift implies $|\nu(M)| = 1$.}
\label{fig:model}
\end{figure*}

Let $g$ be a unitary symmetry under which $\ket{u (\bk)}$ and $\ket{w (\bk)}$ transform as
\bg
U_g (\bk) \ket{a (\bk)} = \ket{a (g\bk)} \, \xi_{a,g} (\bk),
\label{eq:sym_rel}
\eg
for $a= u, w$, where $U_g (\bk)$ is the unitary matrix representing $g$ in the orbital basis, and $\xi_{a,g} (\bk)$ is the corresponding sewing phase.
The following discussion also applies to a normalized reference state $\ket{v (\bk)}$ upon replacing $w$ by $v$.
At a $g$-invariant high-symmetry momentum (HSM) $\bk_* = g \bk_*$ modulo $\bb G$, let $\bb G_{g, \bk_*}$ satisfy $\bk_* = g \bk_* + \bb G_{g, \bk_*}$.
In the periodic gauge, the symmetry operator at $\bk_*$ is $V(\bb G_{g,\bk_*})^{-1} U_g (\bk_*)$~\cite{alexandradinata2014wilson,supple}.
Indeed, Eq.~\eqref{eq:sym_rel} becomes $V(\bb G_{g,\bk_*})^{-1} U_g (\bk_*) \ket{a (\bk_*)} = \xi_{a,g} (\bk_*) \ket{a (\bk_*)}$.
Thus, when $\ket{a (\bk_*)} \ne0$, $\xi_{a,g} (\bk_*)$ is the symmetry eigenvalue.
For a reference state $\ket{w (\bk)}$ that vanishes at $\bk_*$, $\xi_{w,g}(\bk)$ can be taken to be smooth and well defined at $\bk_*$~\cite{supple}.

The overlap then satisfies
\bg
\zeta_w (g\bk) = \xi_{w,g} (\bk) \, \xi_{u,g} (\bk)^{-1} \, \zeta_w (\bk).
\label{eq:zeta_sym}
\eg
If both states defining the overlap are nonzero at $\bk_*$, Eq.~\eqref{eq:zeta_sym} enforces $\zeta_w (\bk_*)=0$ if $\xi_{w,g} (\bk_*) \ne \xi_{u,g} (\bk_*)$.
If $\ket{w (\bk_*)}=0$, the overlap already vanishes, but the sewing phases still constrain its local vorticity, as shown below.

We focus on crystalline symmetries compatible with a nonzero Chern number.
Among the 17 2D wallpaper groups, the nontrivial orientation-preserving groups are $pn$ with $C_n$ rotation symmetry ($n=2,3,4,6$).
A nonzero Chern number further requires broken time-reversal symmetry.
Consider a HSM $\bk_*$ invariant under a $C_m$ rotation, where $C_m$ is a subgroup of $C_n$.
Taking $g = C_m$, we write $\xi_ {a,g} (\bk_*) = e^{\frac{2\pi i}{m} l^{(m)}_{a,\bk_*}}$ where $l^{(m)}_{a,\bk_*} \in \{0,\dots,m-1\}$ denotes the angular momentum modulo $m$.
Denoting $q_\pm = q_x \pm i q_y = q e^{\pm i \phi}$, let the leading-order part of the overlap near $\bk_*$ be
\bg
\zeta_w (\bk_* + \bq) = \sum_{a+b=p} \, c_{a,b} \, q_+^a q_-^b + O(q^{p+1}).
\label{eq:local_expansion}
\eg
The integer vorticity depends on the coefficients of the leading polynomial~\cite{li2024constraints}.
Nevertheless, crystalline symmetry fixes $\nu (\bk_*) \bmod m$.
Each term $c_{a,b} \, q_+^a q_-^b$ has phase dependence $e^{i (a-b) \phi}$ with phase winding $a-b$.
Since Eq.~\eqref{eq:zeta_sym} requires every nonvanishing term to satisfy $a - b = l^{(m)}_{w,\bk_*} - l^{(m)}_{u, \bk_*} \bmod m$~\cite{supple}, it follows that
\bg
\nu(\bk_*) = l^{(m)}_{w,\bk_*} - l^{(m)}_{u, \bk_*} \mod m.
\label{eq:vort_sym}
\eg

Equation~\eqref{eq:vort_sym}, together with the vorticity sum rule in Eq.~\eqref{eq:sum_w}, directly relates the Chern number modulo $n$ to the symmetry-enforced vortices at HSMs.
Generic zeros occur in symmetry-related orbits of $n$ points with the same vorticity, so their total contribution to the Chern number vanishes modulo $n$.
Choosing a reference state with vanishing angular momenta at all HSMs, Eq.~\eqref{eq:vort_sym} gives, for example, in $p4$,
\bg
\mc C_u = l^{(4)}_{u,\Gamma} + l^{(4)}_{u,M} + 2 l^{(2)}_{u,X} \mod 4.
\label{eq:c4_si}
\eg
Here, $\Gamma$, $M$, and $X$ are the standard HSMs of the square BZ.
Thus, the rotation-symmetry indicators in Ref.~\cite{fang2012bulk} follow immediately from vortex counting.

\themecol{\it Model example---}
We illustrate our results using the Lieb-lattice model.
Lieb lattices have been realized in electronic, metamaterial, and cold-atom platforms~\cite{mukherjee2015observation,slot2017experimental,deng2020acoustic,lebrat2026ferrimagnetism}.
The three basis orbitals lie at $\bx_1=(0,0)$, $\bx_2=(1/2,0)$, and $\bx_3=(0,1/2)$ [Fig.~\ref{fig:model}(a)], fixing the matrix $V(\bk)={\rm Diag} (1, e^{i k_x/2}, e^{i k_y/2})$.
The orbital at $\bx_1$ has $C_4$ eigenvalue $+1$, while the orbitals at $\bx_2$ and $\bx_3$ each have $C_2$ eigenvalue $+1$.
Thus, $C_4$ is represented by $U_{C_4}(\bk) = 1 \oplus \sg_x$, where $\sg_x$ exchanges the second and third orbitals.
With the nearest-neighbor hopping amplitude set to unity and spin-orbit coupling $\lambda=0.2$, the Bloch Hamiltonian $H (\bk)$ is given by
\bg
\bpm 0 & 2\cos \frac{k_x}{2} & 2\cos \frac{k_y}{2} \\ 2\cos \frac{k_x}{2} & 0 & 4i \lambda \sin \frac{k_x}{2} \sin \frac{k_y}{2} \\ 2\cos \frac{k_y}{2} & -4i \lambda \sin \frac{k_x}{2} \sin \frac{k_y}{2} & 0 \epm.
\eg

The band structure and symmetry eigenvalues at the HSMs are shown in Fig.~\ref{fig:model}(c).
For the lowest band $\ket{u_1 (\bk)}$, the $C_4$ eigenvalues are $1$ at $\Gamma$ and $-i$ at $M$, and the $C_2$ eigenvalue is $1$ at $X$.
Thus, $(l^{(4)}_{u_1,\Gamma}, l^{(4)}_{u_1,M}, l^{(2)}_{u_1,X})=(0,3,0)$.
For the middle flat band $\ket{u_2 (\bk)}$, $(l^{(4)}_{u_2,\Gamma}, l^{(4)}_{u_2,M}, l^{(2)}_{u_2,X})=(2,0,1)$.
Direct integration of the Berry curvature gives the Chern numbers of these bands as $\mc C_1=-1$ and $\mc C_2=0$, consistent with Eq.~\eqref{eq:c4_si} modulo 4.

For $\ket{u_1 (\bk)}$, we choose the reference state $\ket{w_1 (\bk)}=(1,0,0)^T$.
This is the Fourier transform of the CLS $\ket{w_1 (\bR)} = \ket{\bR, 1}$ [Fig.~\ref{fig:model}(b)].
The overlap $\zeta_{w_1} (\bk) = \brk{w_1 (\bk)}{u_1 (\bk)}$ is thus the projection of $\ket{u_1 (\bk)}$ onto the first orbital.
Since this CLS is centered at $(0,0)$ and has $C_4$ eigenvalue $+1$, $l^{(m)}_{w_1, \bk_*}=0$ at all HSMs.
The angular-momentum mismatch $l^{(4)}_{w_1,M} - l^{(4)}_{u_1,M} = 1 \bmod 4$ enforces a zero of $\zeta_{w_1} (\bk)$ at $M=(\pi,\pi)$ [Fig.~\ref{fig:model}(d)].
Evaluating Eq.~\eqref{eq:vort_def2} gives $\nu(M)=1$, consistent with the local expansion $\zeta_{w_1} (M + \bq) \propto q_+$ in a smooth gauge~\cite{supple}.
This agrees with Eq.~\eqref{eq:vort_sym}.
Since $\mc C_1=-1$, this single zero also saturates the sum rule $\nu_{\rm tot}[\zeta_{w_1}]= - \mc C_1 = 1$.

For the trivial flat band $\ket{u_2(\bk)}$, we instead choose $\ket{w_2 (\bk)} = V(-\bk) \ket{\cm{w}_2 (\bk)}$ with $\ket{\cm{w}_2 (\bk)} = (0, 1 + e^{-i k_y}, - 1 - e^{-i k_x} )^T$.
Here, $\ket{\cm{w}_2 (\bk)}$ is the Fourier transform of the CLS~\cite{supple,hwang2021flat}, $\ket{\bR, 2} + \ket{\bR + \bb a_2, 2} - \ket{\bR, 3} - \ket{\bR + \bb a_1, 3}$.
This state is centered at $(1/2,1/2)$ and carries $C_4$ eigenvalue $-1$, as illustrated in Fig.~\ref{fig:model}(b).
Its sewing phases agree with those of $\ket{u_2 (\bk)}$ at all HSMs, so the overlap $\zeta_{w_2} (\bk) = \brk{w_2 (\bk)}{u_2 (\bk)}$ has no symmetry-enforced zero arising from an angular-momentum mismatch.
Nevertheless, $\zeta_{w_2} (\bk)$ has a single zero $M$, simply because $\ket{w_2 (M)}=0$ [Fig.~\ref{fig:model}(d)].
Since $\mc C_2=0$, the sum rule requires this zero to carry zero vorticity.
Evaluating Eq.~\eqref{eq:vort_def2} gives $\nu(M)=0$.
The local expansion $\zeta_{w_2} (M + \bq) \propto q^2$~\cite{supple} also confirms this result.

In summary, the $\ket{u_2 (\bk)}$ case confirms the sum rule [Eq.~\eqref{eq:sum_w}] even when an overlap zero originates from a zero of the reference state itself, while the $\ket{u_1 (\bk)}$ case illustrates a symmetry-enforced vortex.
Each example has a single overlap zero in the BZ.
Examples with multiple zeros, higher vorticity, and noncontractible zero loop are analyzed in the SM~\cite{supple}.

\themecol{\it Chern topology from local perturbations---}
We now show that the overlap zeros and their vorticities can be inferred from dispersion shifts under a local perturbation.
Consider $H_\eta (\bk) = H (\bk) + \eta \ket{w (\bk)}\bra{w (\bk)}$ and let $H_\eta (\bk) \ket{u (\bk; \eta)} = E_u (\bk; \eta) \ket{u (\bk; \eta)}$ with $\ket{u (\bk; 0)}=\ket{u(\bk)}$.
At an overlap zero $\bk_*$, the perturbation annihilates the unperturbed eigenstate, $H_\eta(\bk_*) \ket{u (\bk_*)} = H (\bk_*) \ket{u (\bk_*)} = E_u (\bk_*) \ket{u (\bk_*)}$, because $\zeta_w (\bk_*) = \brk{w (\bk_*)}{u (\bk_*)} = 0$.
Thus, $\ket{u (\bk_*)}$ remains an exact eigenstate with the same energy for arbitrary $\eta$, so the locations of the overlap zeros can be identified without relying on the weak-perturbation limit.
The Hellmann-Feynman theorem further gives $\der_\eta E_u (\bk; \eta) = |\brk{w (\bk)}{u(\bk; \eta)}|^2$.
For small $\eta$, dispersion shifts by $\delta E_u (\bk) = \eta |\zeta_w (\bk)|^2 + O (\eta^2)$.
Thus, the momentum dependence of the dispersion shift directly probes the absolute square of the overlap.

Near a zero at $\bk_*$, Eq.~\eqref{eq:local_expansion} gives $\delta E_u (\bk_*+\bq) \sim q^{2p}$, with possibly multiple terms at the leading order.
Thus, the leading power of the dispersion shift determines the order $p$ of the zero.
For a zero of order $p$, its possible integer vorticities are $\nu (\bk_*) \in \{-p,-p+2,\dots,p-2,p\}$, while symmetry further requires Eq.~\eqref{eq:vort_sym}.
For an elementary zero, $p=1$, the dispersion shift is quadratic in momentum.
More generally, its leading momentum dependence determines the order $p$ of the overlap zero without measuring the wave-function phase.

For the Lieb model, choosing $\ket{w (\bk)} = (1,0,0)^T$ gives the perturbation $\eta \, {\rm Diag} (1,0,0)$, which is simply an onsite potential on the first orbital [Fig.~\ref{fig:model}(b)].
Such perturbations are accessible in various platforms, for example through local tuning of resonances in metamaterials~\cite{bloch2008many,ozawa2019topological}.
The resulting dispersion changes can be measured in such systems, or probed spectroscopically in electronic bands.
For the lowest Chern band, the overlap zero occurs only at $M$, where the dispersion shift behaves as $\delta E_u (M + \bq) \propto q^2$ [Fig.~\ref{fig:model}(e)].
This implies $p=1$ and hence $|\nu(M)|=1$.
As there is no shift at the other HSMs, these dispersion data alone then give $|\mc C| = 1 \bmod 4$, establishing nontrivial Chern topology without searching the full BZ.
If the $C_4$ eigenvalues are also known, Eq.~\eqref{eq:vort_sym} fixes $\nu(M)=1$.

For systems with multiple zeros, the same analysis applies by accounting for all symmetry-allowed vorticities.
Moreover, the reference state can always be chosen so that $\Gamma$ or a single symmetry orbit of HSMs hosts the symmetry-enforced vorticity, which determines $\mc C_u$ modulo the rotation order at those HSMs~\cite{supple}.

\themecol{\it Discussion---}
Our results connect overlap zeros and their local vorticities to crystalline symmetry and the Chern number, with a direct signature in the dispersion shift under local perturbations.
The construction also extends to other topological invariants.
As an example, the relation to the quantized Berry phase in 1D inversion-symmetric systems is discussed in the SM~\cite{supple}.

For multiband cases, let $\mc U (\bk)$ be a collection of target bands and $\mc V (\bk)$ contain the same number of reference states.
Defining the overlap as $\zeta (\bk) = {\rm Det} [\mc V (\bk)^\dg \mc U (\bk)]$ following the standard determinant construction~\cite{fukui2005chern,marzari2012maximally,li2024constraints}, the results extend by replacing the single-band Chern numbers and angular momenta with their total values.
More generally, our construction may be applicable beyond single-particle Bloch bands, including many-body states defined over a suitable parameter space.

The local structure of an overlap zero also contains information about quantum geometry.
At a zero of $\zeta_v (\bk) = \brk{v (\bk)}{u (\bk)}$ for normalized $\ket{v (\bk)}$, the quantum metrics of the two states obey
\bg
|(\der_i \zeta_v)(\bk_*)| \le \sqrt{g_{u,ii} (\bk_*)} + \sqrt{g_{v,ii} (\bk_*)},
\eg
where $g_{a,ij} (\bk) = {\rm Re} \, \bra{\der_i a (\bk)} \left[ \mathds{1} - \ket{a (\bk)} \bra{a (\bk)} \right] \ket{\der_j a (\bk)}$ is the quantum metric.
For a constant reference state, this reduces to $\frac{1}{2 \eta} \der_i^2 \delta E_u + O(\eta) = |\der_i \zeta_v|^2 \le g_{u,ii}$ at $\bk_*$, so the dispersion shift near the zero provides a lower bound on the quantum metric.
Other momentum directions and unnormalized references can be treated similarly~\cite{supple}.

Allowing the reference state $\ket{w (\bk)}$ to vanish can be important rather than merely a technical generalization.
The simplest example occurs when the full orbital basis forms an elementary band representation whose every gapped single-band split is topological~\cite{bradlyn2017topological,cano2018topology,bradlyn2019disconnected,hwang2026srsi}, as illustrated explicitly in the SM~\cite{supple}.
Any symmetric normalized reference must form such a split, and hence carry a nonzero Chern number~\cite{alexandradinata2018no}, preventing it from being globally smooth.
A reference state allowed to have zeros, however, can retain both symmetry and locality.
CLSs provide a natural realization of such reference states.
Singular flat-band eigenstates have been studied~\cite{rhim2019classification,rhim2020quantum,udagawa2024pinch}.
For a CLS, Ref.~\cite{hwang2021flat} showed that zeros of its Fourier transform imply band touchings when realized as a flat-band eigenstate.
This maps vanishing points of $\ket{w (\bk)}$ to band-touching flat bands and suggests a broader connection between Chern bands and flat-band physics.

Finally, related approaches have characterized Chern number through gauge singularities~\cite{gunawardana2024optimally} and singularities of Bloch eigenstates~\cite{asapanna2026accessing}.
Other formulations use singular or interband Berry connections~\cite{mera2022singular,zhu2024anomalous} and related gauge-covariant quantities in optical and excitonic settings~\cite{xie2024theory,paiva2024shift,chen2026quantum}.
As the Chern number is a property of the line bundle associated with a Bloch band, different connections, such as the Berry connection and the singular connection of Refs.~\cite{mera2022singular,paiva2024shift}, can be used to compute it.
We interpret our formulation in terms of connections and discuss its similarities to and differences from these approaches in the SM~\cite{supple}.
The reference state is an independent input whose symmetry character and real-space form can be chosen for the problem at hand, including smooth unnormalized references that vanish at some momenta.
The resulting zero locations and integer vorticities are gauge independent and connect directly to crystalline symmetry and local perturbative responses.

\themecol{\it Note added---}
Upon completion of this manuscript, we became aware of Ref.~\cite{paiva2026topological}, which relates zeros of wave-function overlaps between two band structures to their relative Chern number and to rotation eigenvalues, and extends the construction to the relative $\Z_2$ invariant of class AII.
Our results agree where they overlap.

\begin{acknowledgments}
The authors thank Bruno Mera, Tobias Holder, and Roni Ilan for helpful discussions.
H.D. acknowledges support from the Engineering and Physical Sciences Research Council (grant number EP/W524323/1).
This work was supported by a UKRI Future Leaders Fellowship MR/Y017331/1.
\end{acknowledgments}

\bibliography{refs.bib}

\let\addcontentsline\oldaddcontentsline
\clearpage

\onecolumngrid
\begin{center}
\textbf{\large Supplemental Material: \\
\ourtitle}
\end{center}

\setcounter{section}{0}
\setcounter{figure}{0}
\setcounter{equation}{0}
\renewcommand{\thefigure}{S\arabic{figure}}
\renewcommand{\theequation}{S\arabic{equation}}
\renewcommand{\thesection}{S\arabic{section}}

\tableofcontents
\hfill \\

\section{Tight-binding notation and compact localized states}

\subsection{Tight-binding Hamiltonian and symmetry}
\label{appsub:notation}
We consider a lattice system with orthonormal basis orbitals $\ket{\bR, \alpha}$, where $\bR$ labels the unit cell and $\alpha = 1, \dots, \Norb$ labels the orbitals within each cell.
The orbital $\ket{\bR, \alpha}$ is located at $\bR + \bx_\alpha$, and $\brk{\bR, \alpha}{\bR', \beta} = \delta_{\bR, \bR'} \, \delta_{\alpha, \beta}$.
A translation-invariant Hamiltonian takes the form
\bg
\hat H = \sum_{\bR, \bR'} \sum_{\alpha, \beta} \, t_{\alpha \beta} (\bR-\bR') \, \ket{\bR, \alpha} \bra{\bR', \beta},
\eg
where $t_{\alpha \beta} (\bR - \bR') = \bra{\bR, \alpha} \hat H \ket{\bR', \beta}$ is the hopping amplitude from $\ket{\bR', \beta}$ to $\ket{\bR, \alpha}$.
Hermiticity requires $t_{\alpha \beta} (\bR - \bR') = t_{\beta \alpha} (\bR' - \bR)^*$.

We define the momentum-space basis by
\bg
\ket{\bk, \alpha} = \frac{1}{\sqrt \Ncell} \sum_\bR \, e^{i \bk \cdot (\bR + \bx_\alpha)} \ket{\bR, \alpha},
\eg
where $\Ncell$ is the total number of unit cells.
In this basis, the momentum-space Hamiltonian is
\bg
H_{\alpha \beta} (\bk) = \bra{\bk, \alpha} \hat H \ket{\bk, \beta}
= \sum_\bR \, t_{\alpha \beta} (\bR) \, e^{-i \bk \cdot (\bR + \bx_\alpha - \bx_\beta)}.
\eg
Its eigenstates and eigenvalues satisfy
\bg
\sum_\beta \, H_{\alpha \beta} (\bk) \, u_{n,\beta} (\bk) = E_n (\bk) \, u_{n, \alpha} (\bk),
\quad
\sum_\alpha \, \left( u_{n,\alpha} (\bk) \right)^* \, u_{m,\alpha} (\bk) = \delta_{nm}.
\eg
We use $\ket{u_n (\bk)}$ to denote the column vector with components $u_{n, \alpha} (\bk)$.
The above equations can then be written in matrix notation as $H(\bk) \ket{u_n (\bk)} = E_n (\bk) \ket{u_n (\bk)}$ and $\brk{u_n (\bk)}{u_m (\bk)} = \delta_{nm}$.

The orbital-embedding matrix is defined as
\bg
V(\bk) = {\rm Diag} \left( e^{i \bk \cdot \bx_1}, \dots, e^{i \bk \cdot \bx_{\Norb}} \right).
\eg
For a reciprocal-lattice vector $\bG$, the basis states obey $\ket{\bk + \bG, \alpha} = e^{i \bG \cdot \bx_\alpha} \ket{\bk, \alpha} = \sum_\beta \, \ket{\bk, \beta} V_{\beta \alpha} (\bG)$.
Thus, the Hamiltonian satisfies
\bg
H(\bk + \bG) = V(\bG)^{-1} H(\bk) V(\bG).
\label{seq:Hk_peri}
\eg
With the same band labeling at $\bk$ and $\bk + \bG$, the eigenstates are therefore related by
\bg
\ket{u_n (\bk + \bG)} = \sum_m \, V (\bG)^{-1} \ket{u_m (\bk)} \, G_{mn} (\bk),
\eg
where $G (\bk)$ is a unitary matrix that mixes states with the same energy.
For an isolated band, this freedom reduces to a $U(1)$ phase.
Throughout this work, we use the periodic gauge~\cite{alexandradinata2014wilson,shiozaki2017topological}, $G_{mn} (\bk) = \delta_{mn}$, in which
\bg
\ket{u_n (\bk + \bG)} = V(\bG)^{-1} \ket{u_n (\bk)}.
\label{seq:peri_gauge}
\eg
A normalized state $\ket{u (\bk)}$ and its projector $P_u (\bk) = \kbr{u (\bk)}{u (\bk)}$ are not strictly periodic in the Brillouin zone (BZ).
In particular, $P_u (\bk + \bG) = V(\bG)^{-1} P_u (\bk) V(\bG)$.
In contrast, the Berry connection and curvature, the overlaps $\zeta_{v,w} (\bk)$, and consequently $\Omega_{v,w} (\bk)$ are periodic functions of $\bk$, as discussed below.
We impose the same periodic-gauge boundary condition $\ket{a (\bk + \bG)} = V (\bG)^{-1} \ket{a (\bk)}$ ($a=v,w)$ on the momentum-space reference states introduced in the main text and below.

We consider a unitary symmetry $g = \{O_g | \bs \delta_g\}$, which acts on position and momentum as $\bb r \to O_g \bb r + \bs \delta_g$ and $\bk \to O_g \bk$, respectively.
Antiunitary symmetries can be treated by including complex conjugation, but we focus on unitary symmetries here.
The action on the basis orbitals is
\bg
\hat g \ket{\bR, \alpha} = \sum_\beta \, \ket{O_g \bR + O_g \bx_\alpha - \bx_\beta + \bs \delta_g, \beta} \, [U_g (\bb 0)]_{\beta \alpha},
\label{seq:sym_basis_orbital}
\eg
where $U_g (\bb 0)$ is a unitary matrix describing the transformation of the orbitals under $g$.
The meaning of the argument $\bb 0$ will become clear below.
The transformed unit-cell label on the right-hand side is a lattice vector $\bR'$ satisfying $\bR' + \bx_\beta = O_g (\bR + \bx_\alpha) + \bs \delta_g$ whenever $[U_g (\bb 0)]_{\beta \alpha} \ne 0$.
In the momentum-space basis, this gives
\bg
\hat g \ket{\bk, \alpha} = \sum_\beta \, \ket{O_g \bk, \beta} \, [U_g (\bk)]_{\beta \alpha},
\quad
U_g (\bk) = e^{-i (O_g \bk) \cdot \bs \delta_g} \, U_g (\bb 0).
\eg
The symmetry condition $\hat g \hat H \hat g^{-1} = \hat H$ then implies $H (O_g \bk) = U_g (\bk) H (\bk) U_g (\bk)^{-1}$.
Thus, the eigenstates at $\bk$ and $O_g \bk$ are related by
\bg
U_g (\bk) \ket{u_n (\bk)} = \sum_m \, \ket{u_m (O_g \bk)} \, B_{g,mn} (\bk),
\label{seq:sewing_multi}
\eg
where $B_{g,mn} (\bk) = \bra{u_m (O_g \bk)} U_g (\bk) \ket{u_n (\bk)}$ is the sewing matrix.
It is unitary and has nonzero entries only when $E_m (O_g \bk) = E_n (\bk)$.
The results below can be extended to multiple bands using the sewing matrix and its determinant.
For simplicity, we focus on a single isolated band $\ket{u (\bk)}$, for which
\bg
U_g (\bk) \ket{u(\bk)} = \ket{u (O_g \bk)} \, \xi_{u,g} (\bk),
\label{seq:sewing}
\eg
with a $U(1)$ sewing phase factor $\xi_{u,g} (\bk)$.

At a $g$-invariant high-symmetry momentum (HSM), let $\bb G_{g, \bk_*}$ be the reciprocal-lattice vector satisfying $\bk_* = O_g \bk_* + \bb G_{g, \bk_*}$.
From the transformation laws of $H (\bk)$ under $g$ and periodicity of $H(\bk)$ [Eq.~\eqref{seq:Hk_peri}], we have $U_g (\bk_*) H (\bk_*) U_g (\bk_*)^{-1} = H(\bk_* - \bb G_{g, \bk_*}) = V(\bb G_{g, \bk_*}) H(\bk_*) V(\bb G_{g, \bk_*})^{-1}$.
Thus, $\left[ V(\bb G_{g, \bk_*})^{-1} U_g (\bk_*), H(\bk_*) \right] = 0$~\cite{alexandradinata2014wilson}.
This means that the symmetry operator at $\bk_*$ is $V (\bb G_{g, \bk_*})^{-1} U_g(\bk_*)$.
In the periodic gauge, the sewing relation also becomes
\bg
V (\bb G_{g, \bk_*})^{-1} U_g (\bk_*) \ket{u (\bk_*)} = \ket{u (\bk_*)} \, \xi_{u,g} (\bk_*).
\label{seq:sym_ev_u}
\eg
Therefore, $\xi_{u,g} (\bk_*)$ is the symmetry eigenvalue.
It is gauge independent because a phase change of $\ket{u(\bk_*)}$ multiplies both sides of Eq.~\eqref{seq:sym_ev_u} by the same phase factor.

\subsection{Compact localized states}
\label{appsub:cls}
We mainly use compact localized states (CLSs) as unnormalized reference states, although more generally $w (\bk)$ may be any smooth global section and need not originate from a CLS.
A CLS can be written as
\bg
\ket{w (\bR)} = \sum_{\bR', \alpha} \, S_\alpha (\bR') \ket{\bR + \bR', \alpha}.
\eg
Compact localization means that there exists a finite radius $R_0$ such that $S_\alpha (\bR) = 0$ for all $\alpha$ whenever $|\bR| > R_0$.
The Fourier transform is
\bg
\ket{W (\bk)} = \frac{1}{\sqrt \Ncell} \sum_\bR \, e^{i \bk \cdot \bR} \ket{w (\bR)}
= \sum_{\bR', \alpha} \, e^{-i \bk \cdot (\bR' + \bx_\alpha)} \, S_\alpha (\bR') \ket{\bk, \alpha}.
\eg
Its components in the momentum-space basis are
\bg
w_\alpha (\bk) = \brk{\bk, \alpha}{W (\bk)}
= e^{-i \bk \cdot \bx_\alpha} \sum_\bR \, e^{-i \bk \cdot \bR} \, S_\alpha (\bR).
\eg
We use $\ket{w (\bk)}$ to denote the column vector with components $w_\alpha (\bk)$.
Although $\ket{W (\bk)}$ and $\ket{w (\bk)}$ are related through the momentum-space basis, we distinguish them because a symmetry $g$ acts on the former by $\hat g$ and on the latter by $U_g (\bk)$, just as on the band eigenstate $\ket{u_n (\bk)}$.
The Fourier transform fixes its boundary condition as
\bg
\ket{w (\bk + \bb G)} = V (\bb G)^{-1} \ket{w(\bk)},
\eg
consistent with the periodic gauge.
Each component is a finite sum of exponential functions of $\bk$.
Therefore, $\ket{w (\bk)}$ is smooth throughout the BZ, although it may vanish at certain momenta.
It is also useful to define $\ket{\cm w (\bk)} = V (\bk) \ket{w (\bk)}$ with components $\cm w_\alpha (\bk) = \sum_\bR \, e^{-i \bk \cdot \bR} \, S_\alpha (\bR)$.
The amplitudes $S_\alpha (\bR)$ can then be read off directly as the coefficients of $e^{-i \bk \cdot \bR}$ in $\cm w_\alpha (\bk)$.

We now consider the symmetry transformation of a CLS centered at $\bR + \bx_w$.
In terms of its center and symmetry representation, it can be treated similarly to a Wannier state, although its Fourier transform may vanish at certain momenta.
We assume that $g$ maps the CLS to itself up to a lattice translation and an overall phase.
This means that $g$ leaves $\bx_w$ invariant up to a lattice translation.
Then, we write
\bg
\hat g \ket{w (\bR)} = \xi_{w,g} (\bb 0) \, \ket{w (O_g \bR + O_g \bx_w - \bx_w + \bs \delta_g)}.
\eg
Here, $O_g \bx_w - \bx_w + \bs \delta_g$ is a lattice vector, and $\xi_{w,g} (\bb 0)$ is a $U(1)$ factor corresponding to a one-dimensional representation of a symmetry group generated by $g$.
(Note that the above equation can be viewed as a simpler version of Eq.~\eqref{seq:sym_basis_orbital} with $\alpha=\beta$.)
Taking the Fourier transform gives
\bg
\hat g \ket{W (\bk)} = \xi_{w,g} (\bk) \ket{W (O_g \bk)},
\quad
\xi_{w,g}(\bk) = \xi_{w,g} (\bb 0) \, e^{-i (O_g \bk) \cdot (O_g \bx_w - \bx_w + \bs \delta_g)}.
\label{seq:w_sewing}
\eg
Equivalently, the component vectors satisfy
\bg
U_g (\bk) \ket{w (\bk)} = \ket{w (O_g \bk)} \, \xi_{w,g} (\bk).
\label{seq:w_sym}
\eg

Now let us consider a $g$-invariant HSM $\bk_*$ and assume that $\ket{w (\bk_*)} \ne 0$.
The same argument leading to Eq.~\eqref{seq:sym_ev_u} shows that $\xi_{w,g} (\bk_*)$ is the eigenvalue of $V (\bb G_{g, \bk_*})^{-1} U_g (\bk_*)$ associated with $\ket{w (\bk_*)}$.
If $\xi_{w,g} (\bk_*)$ is absent from the spectrum of this operator, $\ket{w (\bk_*)}$ must vanish~\cite{hwang2021flat}.

For a set of CLSs that is closed under the symmetry and whose members may mix with one another, we collect their momentum-space component vectors as the columns of a matrix $\mc W (\bk)$.
The symmetry action then takes the form
\bg
U_g (\bk) \mc W (\bk) = \mc W (O_g \bk) B_{\mc W, g} (\bk),
\eg
where the sewing matrix $B_{\mc W,g} (\bk)$ is fixed by the real-space symmetry transformation of the underlying CLS set, including their centers and internal symmetry representation.
For a single CLS, $B_{\mc W,g} (\bk)$ reduces to the sewing phase $\xi_{w,g} (\bk)$ in Eq.~\eqref{seq:w_sym}.
Since $B_{\mc W,g} (\bk)$ is determined from the symmetry action of the CLS set, it is a smooth function of $\bk$ even at momenta where $\mc W (\bk)$ becomes rank deficient, meaning that its columns become linearly dependent.
For a single reference state, this simply corresponds to $\ket{w (\bk)} = 0$.

\section{Local vorticity and the vorticity sum rule}

\subsection{Normalized reference states}
We first consider normalized states $\ket{u (\bk)}$ and $\ket{v (\bk)}$, satisfying $\brk{u (\bk)}{u (\bk)} = \brk{v (\bk)}{v (\bk)} =1$, and derive the sum rule, $\nu_{\rm tot} [\zeta_v] = \mc C_v - \mc C_u$, of the main text.
We comment on the extension to multiple bands at the end of this subsection.
We assume that both $\ket{u (\bk)}$ and $\ket{v (\bk)}$ define smooth projectors throughout the BZ.
The states themselves need only be smooth in local gauge patches, as they can have nonzero Chern numbers $\mc C_{v,u}$.
The overlap between the two states is
\bg
\zeta_v (\bk) = \brk{v (\bk)}{u (\bk)}.
\eg
We first assume that the zeros $\bk_s$ of $\zeta_v (\bk)$ are isolated.
For an extended but contractible zero set, one can use the same construction with any enclosing region whose boundary avoids the zeros.
Noncontractible loops require the BZ boundary identification imposed by the periodic gauge to be treated explicitly and are discussed in Sec.~\ref{appsub:linezero}, with an explicit model example given in Sec.~\ref{appsub:qwz}.
Around each isolated zero, we choose a small disk $D_s$ of radius $\ep$ containing no other zeros and choose smooth gauges for both states on the entire disk.
In this smooth gauge, the vorticity is defined by
\bg
\nu (\bk_s) = \frac{1}{2 \pi} \oint_{\der D_s} d \bk \cdot \der_\bk \arg \zeta_v (\bk)
= \frac{1}{2 \pi} \oint_{\der D_s} d \arg \zeta_v (\bk),
\label{seq:vort_v}
\eg
where $\der D_s$ is the boundary of $D_s$ and is oriented counterclockwise.
We use vector and differential-form notation interchangeably below.
Since $\zeta_v (\bk)$ is smooth and single valued on $\der D_s$ and does not vanish there, its phase changes by an integer multiple of $2\pi$ around the loop.
Thus, $\nu (\bk_s)$ is an integer.
It is also independent of the choice of smooth gauges on $D_s$.
Indeed, two such choices are related by $\ket{a (\bk)} \to e^{i \phi_a (\bk)} \ket{a (\bk)}$ for $a=u,v$, where $\phi_a (\bk)$ can be chosen smooth and single valued on the entire disk.
The resulting change in the vorticity is therefore $\nu' (\bk_s) - \nu (\bk_s) = \frac{1}{2 \pi} \oint_{\der D_s} d(\phi_u-\phi_v)=0$.

We define the Berry connections and curvatures by
\bg
\mc A_a (\bk) = i \brk{a (\bk)}{d a (\bk)},
\quad
\mc F_a (\bk) = d \mc A_a (\bk),
\eg
for $a=u,v$.
In components, for example, $\mc A_a (\bk) = \sum_{i=x,y} \, d k_i \, A_{a,i} (\bk) = \bb A_a (\bk) \cdot d \bk$ where $A_{a,i} (\bk) = i \brk{a (\bk)}{\der_{k_i} a (\bk)}$.
Note that the Berry connection and curvature are periodic in the BZ under the periodic gauge.
Since $V(\bG)$ is independent of $\bk$, $\mc A_a (\bk + \bG) = i \bra{a (\bk)} V(\bG) V(\bG)^{-1} \ket{d a (\bk)} = \mc A_a (\bk)$.
Away from the overlap zeros, we introduce
\bg
\Omega_v (\bk) = d \arg \zeta_v (\bk) + \mc A_u (\bk) - \mc A_v (\bk).
\eg
Under a gauge transformation $\ket{a (\bk)} \to e^{i \phi_a (\bk)} \ket{a (\bk)}$, the overlap transforms as $\zeta_v (\bk) \to e^{i \phi_u (\bk) - i \phi_v (\bk)} \zeta_v (\bk)$ and the connections transform as $\mc A_a (\bk) \to \mc A_a (\bk) - d \phi_a (\bk)$.
These contributions cancel in $\Omega$.
We cover the BZ with gauge patches, each of which is a region where smooth gauges can be chosen for both states.
On overlapping patches, different gauge choices are related by phase factors, called transition functions.
Since $\Omega$ is gauge invariant, it has the same value in either gauge.
Also, since we use the periodic gauge, the overlap is periodic, $\zeta_v (\bk + \bG) = \brk{v (\bk + \bG)}{u (\bk + \bG)} = \bra{v (\bk)} V(\bG) V(\bG)^{-1} \ket{u (\bk)} = \zeta_v (\bk)$.
The Berry connections are also periodic under the periodic gauge.
Finally, within each patch, $d \arg \zeta_v (\bk) = {\rm Im} \, [d \, \zeta_v (\bk) / \zeta_v (\bk)]$ is smooth wherever $\zeta_v (\bk) \ne 0$, and the Berry connections are smooth in the chosen local gauges.
Thus, $\Omega$ is well defined and periodic throughout the BZ away from the overlap zeros.

Let $D_\ep$ denote the union of mutually disjoint small disks $D_s$ around the zeros, $D_\ep = \cup_s D_s$, and define the remaining region of the BZ, $M_\ep = {\rm BZ} \setminus D_\ep$.
We label quantities evaluated in the smooth gauges chosen on $D_s$ and $M_\ep$ by the superscripts $(D_s)$ and $(M_\ep)$, respectively.
In the smooth gauge on $D_s$,
\bg
2 \pi \nu (\bk_s)
= \oint_{\der D_s} d \arg \zeta_v^{(D_s)}
= \oint_{\der D_s} \left( \Omega_v^{(D_s)} - \mc A_u^{(D_s)} + \mc A_v^{(D_s)} \right)
= \oint_{\der D_s} \Omega_v - \int_{D_s} \left( \mc F_u - \mc F_v \right).
\label{seq:vort_local_u}
\eg
The last equality follows from Stokes' theorem applied to the smooth Berry connections on $D_s$.
In the last equality, we have omitted the patch superscripts since $\Omega_v$ and $\mc F_{u,v}$ are gauge invariant.
Defining the total vorticity as $\nu_{\rm tot} [\zeta_v] = \sum_s \, \nu (\bk_s)$, we obtain
\bg
2 \pi \nu_{\rm tot} [\zeta_v] = \oint_{\der D_\ep} \Omega_v - \int_{D_\ep} \left( \mc F_u - \mc F_v \right).
\label{seq:sum_rule_deriv_1}
\eg
On $M_\ep$, $\Omega_v$ is a smooth one-form, so its exterior derivative is well defined.
Evaluating it in the smooth gauge on $M_\ep$ gives $d \Omega_v = \mc F_u - \mc F_v$.
Since $\der M_\ep$ has the opposite orientation to $\der D_\ep$, Stokes' theorem gives $\oint_{\der D_\ep} \Omega_v = - \oint_{\der M_\ep} \Omega_v = - \int_{M_\ep} (\mc F_u - \mc F_v)$.
Hence, Eq.~\eqref{seq:sum_rule_deriv_1} becomes
\bg
2 \pi \nu_{\rm tot} [\zeta_v] = - \int_{M_\ep} \left( \mc F_u - \mc F_v \right) - \int_{D_\ep} \left( \mc F_u - \mc F_v \right)
= - \int_{\rm BZ} \left( \mc F_u - \mc F_v \right)
= 2 \pi (\mc C_v - \mc C_u),
\eg
where we used $C_{u,v} = \frac{1}{2 \pi} \int_{\rm BZ} \mc F_{u,v}$.
The sum rule is thus given by
\bg
\nu_{\rm tot} [\zeta_v] = \mc C_v - \mc C_u.
\label{seq:sum_rule_u}
\eg

The same construction extends directly to the multiband case.
Consider two sets of $N$ states and write $\mc U (\bk)$ and $\mc V (\bk)$ for the $\Norb \times N$ matrices whose columns span the target and reference band subspaces, respectively.
Explicitly, $\mc U_{\alpha n} (\bk) = u_{n, \alpha} (\bk)$ and
$\mc V_{\alpha n} (\bk) = v_{n, \alpha} (\bk)$, where $\alpha = 1, \dots, \Norb$ labels the orbital basis and $n = 1,\dots,N$ labels the states within each subspace.
Then, one replaces $\zeta_v (\bk) = \brk{v (\bk)}{u (\bk)}$ by $\zeta_{\mc V} (\bk) = {\rm Det} [\mc V (\bk)^\dg \mc U (\bk)]$ and the Berry connections by $\mc A_{\mc U} (\bk) = i {\rm Tr} \, [\mc U (\bk)^\dg d \mc U (\bk)]$ and similarly for $\mc A_{\mc V} (\bk)$.
Thus, $\Omega_{\mc V} = d \arg \zeta_{\mc V} + \mc A_{\mc U} - \mc A_{\mc V}$, and the same argument gives
\bg
\nu_{\rm tot} [\zeta_{\mc V}] = \mc C_{\mc V} - \mc C_{\mc U},
\eg
where $\mc C_{\mc U, \mc V}$ are the total Chern numbers of the corresponding multiband subspaces.
The unnormalized case considered below generalizes in the same way.

We now turn to the local vorticity itself and derive an expression that can be evaluated numerically.
Since the projectors of $u$ and $v$ are smooth, both $\mc F_u$ and $\mc F_v$ are smooth, and the surface integral in Eq.~\eqref{seq:vort_local_u} vanishes as $\ep \to 0$.
Therefore, the local vorticity $\nu(\bk_s)$ can be expressed as
\bg
\nu (\bk_s) = \lim_{\ep \to 0} \frac{1}{2 \pi} \oint_{\der D_s} \Omega_v.
\label{seq:local_formula}
\eg
To evaluate Eq.~\eqref{seq:local_formula} numerically, we discretize $\der D_s$ into a sequence of momenta $\bk_i$ ($i=1,\dots,N$), with $\bk_N = \bk_1$, ordered along the positive orientation of $\der D_s$.
For sufficiently small steps $\Delta \bk_i = \bk_{i+1} - \bk_i$, the line integral of $\Omega_v$ along each segment can be approximated as
\bg
\int_{\bk_i}^{\bk_{i+1}} \Omega_v
\simeq \arg \zeta_v (\bk_{i+1}) - \arg \zeta_v (\bk_i)
+ \arg \brk{u (\bk_{i+1})}{u (\bk_i)} - \arg \brk{v (\bk_{i+1})}{v (\bk_i)}.
\label{seq:omega_link1}
\eg
This can be seen from $\brk{u (\bk+\Delta \bk)}{u (\bk)} \simeq e^{i \int_{\bk}^{\bk + \Delta \bk} {\mc A}_u}$.
The phase differences in Eq.~\eqref{seq:omega_link1} can be combined into the phase of a single complex quantity.
For each oriented segment connecting two neighboring points $\bk_i$ and $\bk_{i+1}$, we define the gauge-invariant link phase
\bg
\td \Omega_{v,i} = {\rm Im} \log \left[ \frac{\zeta_v (\bk_{i+1})}{\zeta_v (\bk_i)} \frac{\brk{u (\bk_{i+1})}{u (\bk_i)}}{\brk{v (\bk_{i+1})}{v (\bk_i)}} \right],
\eg
where the logarithm is taken on a fixed branch, for example, $\arg (z \ne 0) \in (-\pi,\pi]$ for $z \in \C$.
Here, {\it link} refers to the oriented segment $\bk_i \to \bk_{i+1}$ of the discretized contour.
The phase factors associated with gauge transformations of $u$ and $v$ cancel within each link, so $\td \Omega_{v,i}$ can be evaluated without choosing a specific gauge along the contour.
The discretization is chosen sufficiently fine such that the phase accumulated on each link does not cross the branch cut of the logarithm.
Therefore, for a sufficiently fine discretization, $\oint_{\der D_s} \Omega_v \simeq \sum_{i=1}^{N-1} \td \Omega_{v,i}$ and
\bg
\nu(\bk_s) = \lim_{\ep \to 0} \frac{1}{2 \pi} \sum_{i=1}^{N-1} \td \Omega_{v,i}.
\eg

The limit $\ep \to 0$ is not necessary if one keeps the Berry-curvature contribution in Eq.~\eqref{seq:vort_local_u}, as long as $D_s$ contains no other zero and no zero lies on $\der D_s$.
For a finite contour $\der D_s$, the local vorticity can instead be evaluated as
\bg
\nu (\bk_s) = \frac{1}{2 \pi} \left[ \sum_{i=1}^{N-1} \td \Omega_{v,i} - \int_{D_s} (\mc F_u - \mc F_v) \right].
\label{seq:vort_v_numeric}
\eg

\subsection{Unnormalized reference states}
We now consider an unnormalized reference state $\ket{w (\bk)}$.
The derivation parallels that for a normalized reference state $\ket{v (\bk)}$.
We take $\ket{w (\bk)}$ to be a fixed smooth function of $\bk$, subject to the periodic-gauge boundary condition specified by $V(\bG)$ [Eq.~\eqref{seq:peri_gauge}], which may vanish at some momenta, and define
\bg
\zeta_w (\bk) = \brk{w (\bk)}{u (\bk)}.
\eg
The zeros of $\zeta_w (\bk)$ can have two different origins.
A zero of $\ket{w (\bk)}$ necessarily gives a zero of $\zeta_w (\bk)$, while additional zeros can occur when $\ket{w (\bk)}$ is nonzero but orthogonal to $\ket{u (\bk)}$.
We do not need to distinguish these cases below and treat all zeros of $\zeta_w (\bk)$ on the same footing.

Since $\ket{w (\bk)}$ is fixed and carries no additional gauge freedom, we define a gauge invariant one-form
\bg
\Omega_w (\bk) = d \arg \zeta_w (\bk) + \mc A_u (\bk).
\eg
Away from the zeros of $\zeta_w$, $\Omega_w$ is smooth and satisfies $d\Omega_w=\mc F_u$.
Consider an isolated zero at $\bk_s$ and a small disk $D_s$ enclosing it.
In a smooth gauge on $D_s$, its local vorticity is
\bg
2 \pi \nu (\bk_s) = \oint_{\der D_s} d \arg \zeta_w
= \oint_{\der D_s} \Omega_w - \int_{D_s} \mc F_u
= \lim_{\ep \to 0} \oint_{\der D_s} \Omega_w.
\label{seq:local_formula_w}
\eg
Summing over all zeros and repeating the argument used for the normalized reference state gives $2 \pi \nu_{\rm tot} [\zeta_w] = - \int_{\rm BZ} \mc F_u$,
and hence
\bg
\nu_{\rm tot} [\zeta_w] = - \mc C_u.
\eg

We next give the discretized expression for the local vorticity of $\zeta_w (\bk)$.
Discretizing $\der D_s$ into momenta $\bk_i$ as above, a gauge-invariant link phase can be defined as
\bg
\td \Omega_{w,i} = {\rm Im} \log \left[ \frac{\zeta_w (\bk_{i+1})}{\zeta_w (\bk_i)} \brk{u (\bk_{i+1})}{u (\bk_i)} \right],
\label{seq:tdOm_w}
\eg
and, $\oint_{\der D_s} \Omega_w \simeq \sum_{i=1}^{N-1} \td \Omega_{w,i}$ for a sufficiently fine discretization.
In the limit $\ep \to 0$, the local vorticity is given by
\bg
\nu (\bk_s) = \lim_{\ep \to 0} \frac{1}{2 \pi} \sum_{i=1}^{N-1} \td \Omega_{w,i}.
\eg
For a finite disk $D_s$ containing no other zero and with no zero on its boundary,
\bg
\nu (\bk_s) = \frac{1}{2 \pi} \left[ \sum_{i=1}^{N-1} \td \Omega_{w,i} - \int_{D_s} \mc F_u \right].
\label{seq:vort_w_numeric}
\eg

\subsection{Extended zeros}
\label{appsub:linezero}
For a contractible extended zero set, one may simply surround it by a closed contour and define its total vorticity from the winding along that contour, exactly as for an isolated zero.
The construction also applies when the zero set is noncontractible.
For concreteness, we consider the unnormalized case and suppose that $\zeta_w (\bk)$ vanishes along the noncontractible line $L = \{ \bk | k_y = 0 \}$.
The normalized case can be treated in the same way by replacing $\Omega_w$ and $\mc F_u$ with $\Omega_v$ and $\mc F_u-\mc F_v$, respectively.
We surround $L$ by a narrow strip $D_\ep = \{ \bk | |k_y| < \ep \}$, chosen such that no other zeros are contained in $D_\ep$ or lie on its boundary.
The boundary $\der D_\ep$ consists of the two noncontractible lines $k_y = \pm \ep$, with opposite orientations.
The vorticity associated with the line zero is then
\bg
\nu(L) = \frac{1}{2 \pi} \oint_{\der D_\ep} d \, {\rm arg} \, \zeta_w
= \frac{1}{2 \pi} \left[ \oint_{\der D_\ep} \Omega_w - \int_{D_\ep} \mc F_u \right].
\eg
(Although the strip $D_\ep$ is noncontractible, a smooth gauge can be chosen on $D_\ep$, so $\oint_{\der D_\ep} \mc A_u = \int_{D_\ep} \mc F_u$.)
This has the same form as the expression for an isolated zero, except that the enclosing region is now a strip and its boundary remains extended as $\ep \to 0$.

To make the boundary integral explicit, we represent the strip in a domain with $0 \le k_x \le G_x = 2 \pi$.
The two edges at $k_x=0$ and $k_x=G_x$ are identified by the periodic-gauge boundary condition, $\ket{u(G_x,k_y)} = V(G_x)^{-1} \ket{u(0,k_y)}$.
Consider, for example, the discretized integral along $k_y=+\ep$, with $\bk_1 = (0,\ep)$ and $\bk_N = (G_x, \ep)$.
If the states at the two endpoints are evaluated independently, the gauge-invariant link variable [Eq.~\eqref{seq:tdOm_w}] can be used directly.
Alternatively, the BZ boundary identification can be made explicit in the last link as
\bg
\td \Omega_{w,N-1}
= {\rm Im} \log \left[ \frac{\zeta_w (\bk_N)}{\zeta_w (\bk_{N-1})} \brk{u (\bk_N)}{u (\bk_{N-1})} \right]
= {\rm Im} \log \left[ \frac{\zeta_w (\bk_1)}{\zeta_w (\bk_{N-1})} \bra{u (\bk_1)} V(\bG) \ket{u (\bk_{N-1})} \right].
\eg
Here we used $\ket{u (G_x,\ep)} = V((G_x,0))^{-1} \ket{u (0,\ep)}$ and $\zeta_w (G_x,\ep) = \zeta_w (0,\ep)$.
Applying the same boundary identification to both components of $\der D_\ep$, we obtain
\bg
2 \pi \nu(L) = \int_0^{G_x} dk_x
\left[ \Omega_{w,x} (k_x,-\ep) - \Omega_{w,x} (k_x,+\ep) \right] - \int_{D_\ep} \mc F_u,
\eg
where $\Omega_{w,x}$ denotes the $d k_x$ component of the one-form $\Omega_w = \Omega_{w,x} \, dk_x + \Omega_{w,y} \, dk_y$.
Since $\mc F_u$ is smooth, the area of $D_\ep$ vanishes linearly with $\ep$.
Thus,
\bg
\nu (L) = \lim_{\ep \to 0} \frac{1}{2 \pi} \int_0^{G_x} d k_x \, \left[ \Omega_{w,x} (k_x, -\ep) - \Omega_{w,x} (k_x, +\ep) \right].
\label{seq:vort_line}
\eg
Unlike the case of a point zero, the Berry-curvature contribution for a line zero vanishes only linearly with $\ep$.
Indeed, the area of a disk of radius $\ep$ scales as $O(\ep^2)$, whereas the area of a strip surrounding a noncontractible line scales as $O(\ep)$.
Consequently, Eq.~\eqref{seq:vort_line} converges more slowly for a line zero.
For numerical calculations at finite $\ep$, it is preferable to retain the Berry-curvature term.
When additional point zeros are present or when several extended zero sets intersect, the same construction can be applied by choosing a region whose boundary avoids all zeros and consistently identifying the BZ seams by the matrix $V(\bG)$.

When line zeros are present, their vorticities $\nu(L)$ should be included together with the vorticities of any isolated zeros in the total vorticity.
Once all zero sets are enclosed by regions with well-defined boundaries, the derivation of the sum rule follows exactly as above.
The boundary integrals are converted to Berry-curvature integrals over the complementary regions, and together all regions cover the BZ.
Therefore, the same vorticity sum rule remains valid, giving $\nu_{\rm tot} [\zeta_w] = -\mc C_u$ for the unnormalized reference-state case considered here.

\section{Rotation symmetry and vorticity}

\subsection{Normalized reference states}
We consider an $n$-fold rotation $C_n$ and a $C_n$-invariant high-symmetry momentum (HSM) $\bk_*$, with $C_n \bk_*=\bk_*$, and use local momentum coordinates centered at $\bk_*$.
At $\bk_*$, a normalized state $\ket{a (\bk_*)}$ ($a=u,v$) can be chosen as an eigenstate of the rotation operator,
\bg
V (\bG_{C_n, \bk_*} )^{-1} U_{C_n} (\bk_*) \ket{a (\bk_*)} = \xi_{a, C_n} (\bk_*) \ket{a (\bk_*)},
\eg
where
\bg
\xi_{a,C_n} (\bk_*) = e^{\frac{2 \pi i}{n} l^{(n)}_{a, \bk_*}}, \quad
l^{(n)}_{a, \bk_*} \in \{0, \dots, n-1\}.
\label{seq:ang_l}
\eg
Throughout this work, the superscript $(n)$ on $l^{(n)}_{a, \bk_*}$ indicates that the angular momentum is defined modulo $n$.
Under the rotation, the overlap satisfies
\bg
\zeta_v (C_n \bk) = \frac{\xi_{v, C_n} (\bk)}
{\xi_{u, C_n} (\bk)} \zeta_v (\bk),
\label{seq:rot_sewing}
\eg
where we used Eq.~\eqref{seq:sewing}.
Writing the Berry-connection one-form as $\mc A_a (\bk) = \bb A_a (\bk) \cdot d\bk$, its transformation under the rotation is $\mc A_a (C_n \bk) = \mc A_a (\bk) + d \, {\rm arg} \, \xi_{a, C_n} (\bk)$, where $\mc A_a (C_n \bk) = \bb A_a (C_n \bk) \cdot d(C_n \bk)$.
Equivalently, in vector notation, $\bb A_a (C_n \bk) \cdot d(C_n \bk) = \bb A_a (\bk) \cdot d \bk + \der_\bk \, {\rm arg} \, \xi_{a, C_n} (\bk) \cdot d \bk$.
Together with Eq.~\eqref{seq:rot_sewing}, the sewing-phase contributions cancel in $\Omega_v (\bk) = \bb \Omega_v (\bk) \cdot d \bk$, giving
\bg
\Omega_v (C_n \bk) = \Omega_v (\bk),
\quad
\bb \Omega_v (C_n\bk) \cdot d(C_n \bk) = \Omega_v (\bk) \cdot d \bk,
\eg

Now, consider a small circle $\der D_\ep$ of radius $\ep$ centered at $\bk_*$ and choose an oriented arc $\gamma_\ep$ connecting a point $\bk_* + \bq$, with $|\bq| = \ep$, to its rotated counterpart $\bk_* + C_n \bq$.
The remaining parts of the circle are obtained by successive $C_n$ rotations of this arc.
Since the line integral of $\Omega_v$ is unchanged under the rotation, $\oint_{\der D_\ep} \Omega_v =
n \int_{\gamma_\ep} \Omega_v$.
We choose smooth gauges for both states on $D_\ep$.
Since the Berry connections are smooth, their contribution along $\gamma_\ep$, whose length is $O(\ep)$, vanishes as $\ep \to 0$.
Using the relation above, we obtain
\bg
\frac{2 \pi}{n} \nu (\bk_*)
= \lim_{\ep \to 0} \int_{\gamma_\ep} \Omega_v
= \lim_{\ep \to 0} \int_{\bk_*+ \bq}^{\bk_* + C_n \bq} d \, {\rm arg} \, \zeta_v
= \lim_{\ep \to 0} \big[ {\rm arg} \, \zeta_v (\bk_* + C_n \bq) - {\rm arg} \, \zeta_v (\bk_* + \bq) + 2\pi \Z \big].
\eg
Using the rotation constraint in Eq.~\eqref{seq:rot_sewing}, $\zeta_v (\bk_* + C_n \bq) / \zeta_v (\bk_* + \bq) = \xi_{v, C_n}(\bk_* + \bq) / \xi_{u, C_n}(\bk_* + \bq)$, and hence
\bg
\frac{2\pi}{n} \nu (\bk_*) = \lim_{\ep \to 0} {\rm arg} \left[ \frac{\xi_{v, C_n} (\bk_* + \bq)}{\xi_{u, C_n} (\bk_* + \bq)} \right] \mod 2 \pi.
\eg
Taking $\bq \to 0$, $\xi_{a, C_n} (\bk_* + \bq) \to e^{\frac{2 \pi i}{n} l^{(n)}_{a, \bk_*}}$ and thus
\bg
\nu(\bk_*) = l^{(n)}_{v, \bk_*} - l^{(n)}_{u, \bk_*} \mod n.
\label{seq:vort_rot_v}
\eg
Note that we used smooth gauges in which the Berry connections and the sewing phases $\xi_{a, C_n}$ are smooth.
The final result, however, is gauge invariant.

Equation~\eqref{seq:vort_rot_v} also implies that a mismatch between $l^{(n)}_{v, \bk_*}$ and $l^{(n)}_{u, \bk_*}$ enforces an overlap zero at $\bk_*$, with vorticity given by their difference modulo $n$.
The same symmetry constraint can also be seen directly from the local form of the overlap.
Following the notation of the main text, we write $q_\pm = q_x \pm i q_y = q e^{\pm i \phi}$, where $q = |\bq|$ and $\phi$ is the polar angle of $\bq$.
Choosing smooth local gauges around $\bk_*$, the overlap is nonsingular there and can be expanded regularly in powers of $\bq$.
 We write its leading nonzero term as $\zeta_v (\bk_* + \bq) = \sum_{a+b=p} \, c_{a,b} q_+^a q_-^b + O(q^{p+1})$.
Under $C_n$, $q_\pm \to e^{\pm 2 \pi i/n} q_\pm$.
On the other hand, Eq.~\eqref{seq:rot_sewing} gives, to leading order in $\bq$,
\bg
\zeta_v (\bk_* + C_n \bq) = e^{\frac{2\pi i}{n} \left( l^{(n)}_{v, \bk_*} - l^{(n)}_{u, \bk_*} \right)} \zeta_v (\bk_* + \bq) + O(q^{p+1}).
\eg
Therefore, every nonzero coefficient $c_{a,b}$ in the leading-order term satisfies
\bg
a-b = l^{(n)}_{v, \bk_*} - l^{(n)}_{u, \bk_*}
\mod n.
\eg
Thus, rotation symmetry constrains the allowed leading-order form of the overlap near $\bk_*$, in accordance with Eq.~\eqref{seq:vort_rot_v}.

The multiband extension proceeds by replacing the scalar sewing phases with sewing matrices.
For $\zeta_{\mc V} (\bk) = {\rm Det} [\mc V^\dg (\bk) \mc U (\bk)]$, one finds $\zeta_{\mc V} (C_n \bk) = \frac{{\rm Det} B_{\mc V} (\bk)}{{\rm Det} B_{\mc U} (\bk)} \zeta_{\mc V} (\bk)$, where $B_{\mc U} (\bk)$ and $B_{\mc V} (\bk)$ are the $C_n$ sewing matrices of the target and reference subspaces, respectively [Eq.~\eqref{seq:sewing_multi}].
Defining the total angular momenta through $\Det B_{A= \mc U, \mc V} (\bk_*) = e^{\frac{2 \pi i}{n} l^{(n)}_{A, \bk_*}}$, one obtains $\nu (\bk_*) = l^{(n)}_{\mc V, \bk_*} - l^{(n)}_{\mc U, \bk_*} \bmod n$.
The same construction applies to an unnormalized multiband reference $\mc W (\bk)$ as discussed below.

\subsection{Unnormalized reference states}
\label{appsub:rot_condition_w}
We now consider an unnormalized reference state $\ket{w (\bk)}$.
The derivation closely parallels that for a normalized reference state, except that $\ket{w (\bk_*)}$ itself may vanish at the rotation-invariant momentum.
The symmetry relation for the reference state can be written as [Eq.~\eqref{seq:w_sym}]
\bg
U_{C_n} (\bk) \ket{w (\bk)} = \xi_{w, C_n} (\bk) \ket{w (C_n \bk)}.
\eg
For the pure rotation considered here, the sewing phase takes the form [Eq.~\eqref{seq:w_sewing}]
\bg
\xi_{w, C_n} (\bk) = \xi_{w, C_n} (\bb 0) e^{-i (C_n \bk) \cdot (C_n \bx_w - \bx_w)}.
\label{seq:w_sewing_rot}
\eg
If $\ket{w (\bk_*)}$ is nonzero, $\xi_{w, C_n} (\bk_*)$ is its symmetry eigenvalue.
Even when $\ket{w (\bk_*)}=0$, however, the sewing phase $\xi_{w, C_n} (\bk)$ remains well defined around $\bk_*$ through Eq.~\eqref{seq:w_sewing_rot}.
We therefore use the same angular-momentum label modulo $n$ for its value at $\bk_*$ and write
\bg
\xi_{w, C_n} (\bk_*) = e^{\frac{2 \pi i}{n} l^{(n)}_{w, \bk_*}},
\quad
l^{(n)}_{w, \bk_*} \in \{0, \dots, n-1 \}.
\eg
Using the symmetry relations for $\ket{u (\bk)}$ and $\ket{w (\bk)}$, the overlap satisfies
\bg
\zeta_w (C_n \bk) = \frac{\xi_{w, C_n} (\bk)}{\xi_{u, C_n} (\bk)} \zeta_w (\bk).
\eg
Combining this relation with the transformation of $\mc A_u$, we obtain
\bg
\Omega_w (C_n \bk) = \Omega_w (\bk) - (\bx_w - C_n^{-1} \bx_w) \cdot d \bk.
\eg

We use the same circle $\der D_\ep$ and oriented arc $\gamma_\ep$ as in the normalized case.
Unlike $\Omega_v$, the integral of $\Omega_w$ is not exactly identical on symmetry-related arcs at finite $\ep$.
Nevertheless, the difference between the integrals on two symmetry-related arcs is $O(\ep)$:
\bg
\int_{\gamma_\ep} (\bx_w - C_n^{-1} \bx_w) \cdot d \bk = O(\ep)
\label{seq:small_arc}
\eg
Consequently,
\bg
\lim_{\ep \to 0} \oint_{\der D_\ep} \Omega_w = n \lim_{\ep \to 0} \int_{\gamma_\ep} \Omega_w.
\eg
Using this together with the small-circle expression for the local vorticity [Eq.~\eqref{seq:local_formula_w}], and choosing a smooth gauge for $\ket{u (\bk)}$ on $D_\ep$, the Berry-connection contribution along $\gamma_\ep$ vanishes as $\ep \to 0$.
Hence,
\bg
\frac{2 \pi}{n} \nu (\bk_*) = \lim_{\ep \to 0} \int_{\gamma_\ep} \Omega_w
= \lim_{\ep \to 0} \left[ {\rm arg} \, \zeta_w (\bk_* + C_n \bq) - {\rm arg} \, \zeta_w (\bk_* + \bq) + 2 \pi \Z \right].
\eg
Using the symmetry constraint on $\zeta_w$ and taking $\bq \to 0$, we obtain
\bg
\nu (\bk_*) = l^{(n)}_{w, \bk_*} - l^{(n)}_{u, \bk_*}
\mod n.
\label{seq:vort_rot_w}
\eg
When $\ket{w (\bk_*)} = 0$, the usual eigenvalue interpretation of $\xi_{w, C_n} (\bk_*)$ is no longer available.
Nevertheless, $\xi_{w, C_n} (\bk)$ remains well defined and constrains the behavior of $\ket{w (\bk)}$ around $\bk_*$.
Thus, although the zero of $\zeta_w$ is inherited from a zero of the reference state, its local vorticity is constrained by rotation symmetry in the same way as for a symmetry-enforced overlap zero.
The local expansion of $\zeta_w (\bk)$ therefore follows in the same way as in the normalized case.

The multiband extension proceeds in the same way as in the normalized case.
For a set of unnormalized reference states $\mc W (\bk)$, we define $\zeta_{\mc W} (\bk) = {\rm Det} [\mc W^\dg (\bk) \mc U (\bk)]$ and use its $C_n$ sewing matrix $B_{\mc W} (\bk)$.
For the CLS reference states considered here, $B_{\mc W} (\bk)$ is fixed by the real-space symmetry action of the underlying CLS set and is a smooth function of $\bk$, as discussed in Sec.~\ref{appsub:cls}.
Therefore, even when $\mc W (\bk_*)$ is rank deficient, ${\rm Det} B_{\mc W} (\bk_*)$ remains well defined.
Defining ${\rm Det} B_{\mc W} (\bk_*) = e^{\frac{2 \pi i}{n}l^{(n)}_{\mc W, \bk_*}}$, one obtains $\nu (\bk_*) = l^{(n)}_{\mc W, \bk_*} - l^{(n)}_{\mc U, \bk_*} \bmod n$.
The multiband counterpart of the correction in Eq.~\eqref{seq:small_arc} is $i \int_{\gamma_\epsilon} {\rm Tr} \left[ B_{\mc W}^\dg (\bk) d B_{\mc W} (\bk) \right] = O(\ep)$, which follows from the smoothness of $B_{\mc W} (\bk)$ near $\bk_*$.
More generally, the same conclusion holds whenever a smooth sewing matrix for $\mc W (\bk)$ is specified.

\subsection{Concentrating angular-momentum mismatch}
The symmetry-enforced vorticity can be concentrated at a single symmetry orbit of HSMs by a suitable choice of reference state.
Here, by an orbit we mean a set of HSMs related by $C_n$, such as $\{\Gamma\}$, $\{M\}$, and $\{X,Y\}$ for $C_4$, or $\{\Gamma\}$, $\{K,K'\}$, and $\{M,M',M''\}$ for $C_6$.
Here, we use the standard notation for HSMs in $C_n$-symmetric systems.
For a $C_n$-symmetric band or subspace $A$, ${\rm Det} B_A (\bk_*) = e^{\frac{2\pi i}{m} l^{(m)}_{A, \bk_*}}$ as defined above, where $C_m$ is the little group at $\bk_*$ and $l^{(m)}_{A, \bk_*} \in \{0,1,\dots,m-1\}$.
Note that $m$ divides $n$.
The standard rotation-eigenvalue expressions for the Chern number modulo $n$~\cite{fang2012bulk} can be written in terms of these angular momenta as
\ba
C_2 :& \, \, \mc C_A = l^{(2)}_{A, \Gamma} + l^{(2)}_{A,X} +l^{(2)}_{A,Y} + l^{(2)}_{A,M} \mod 2,
\quad
&& C_3 : \, \, \mc C_A = l^{(3)}_{A,\Gamma} + l^{(3)}_{A,K} + l^{(3)}_{A,K'} \mod 3,
\nn
C_4 :& \, \, \mc C_A = l^{(4)}_{A,\Gamma} + l^{(4)}_{A,M} + 2l^{(2)}_{A,X} \mod 4,
\quad
&& C_6 : \, \, \mc C_A = l^{(6)}_{A,\Gamma} + 2 l^{(3)}_{A,K} + 3 l^{(2)}_{A,M} \mod 6.
\label{seq:modn_Chern}
\ea
These relations are also recovered directly from our vorticity sum rule.

Let us first summarize the result.
Consider an orbit of $p=n/m$ HSMs related by $C_n$, whose little group is $C_m$.
If the angular-momentum mismatch is confined to this orbit, the $p$ symmetry-related points carry the same vorticity $\nu$ modulo $m$, giving $\mc C_u = p \nu$ modulo $n$.
Such a concentration is therefore possible only when $\mc C_u$ is a multiple of $p$ modulo $n$.
Conversely, for the $C_n$ symmetries considered here, we show below that whenever $\mc C_u \in p \Z \bmod n$, a reference state can be chosen so that the mod-$n$ total vorticity is concentrated on an orbit of $p$ HSMs.

For our purposes, we use an unnormalized reference state $\ket{w (\bk)}$, although the same discussion applies to normalized reference states.
Although $\ket{w (\bk)}$ may vanish and therefore need not define a Chern band, its angular momenta are determined by the real-space symmetry transformation of the underlying CLS, in direct analogy with a Wannier representation.
Consequently, substituting $l^{(m)}_{w, \bk_*}$ into Eq.~\eqref{seq:modn_Chern} always gives zero modulo $n$.
We use only this constraint below, without assigning a Chern number to $\ket{w (\bk)}$.
For a multiband reference, the same statement applies to the total angular momenta defined through ${\rm Det} B_W (\bk)$.

\paragraph{$C_2$ symmetry:}
For $C_2$, consider a single CLS centered at a Wyckoff position with unit multiplicity.
Such position can be represented as $\bx_w = (a/2, b/2)$ $a,b \in \{0,1\}$.
At each position, the CLS can have angular momentum $l \in \{0,1\}$ since the site-symmetry group of $\bx_w$ is isomorphic to $C_2$.
From Eq.~\eqref{seq:w_sewing}, its angular momenta are given by
\bg
\left( l^{(2)}_{w, \Gamma}, l^{(2)}_{w, X}, l^{(2)}_{w, Y}, l^{(2)}_{w, M} \right)
= \left( l, l+a, l+b, l+a+b \right)
\mod 2.
\eg
In fact, the above equation gives the general solution of Eq.~\eqref{seq:modn_Chern} with zero on the right-hand side for $C_2$.
To concentrate the angular-momentum mismatch, which enforces nonzero vorticity modulo 2, at $\Gamma$, we choose $l,a,b$ such that
\bg
l+a = l^{(2)}_{u, X}, \quad
l+b = l^{(2)}_{u, Y},\quad
l+a+b = l^{(2)}_{u, M}.
\eg
These equations always have a solution.
They give $l^{(2)}_{w, \Gamma} = l^{(2)}_{u, X} + l^{(2)}_{u, Y} + l^{(2)}_{u, M}$ mod 2, and hence
\bg
l^{(2)}_{u, \Gamma} - l^{(2)}_{w, \Gamma}
= l^{(2)}_{u, \Gamma} + l^{(2)}_{u, X} + l^{(2)}_{u, Y} + l^{(2)}_{u, M}
= \mc C_u \mod 2.
\eg
Thus the symmetry-enforced mismatch is concentrated at $\Gamma$.
By matching a different set of three HSMs, the mismatch can instead be placed at $X$, $Y$, or $M$.

\paragraph{$C_3$ symmetry:}
For $C_3$, the three multiplicity-one Wyckoff positions give
\bg
\left( l^{(3)}_{w, \Gamma}, l^{(3)}_{w, K}, l^{(3)}_{w, K'} \right)
=
\left( l, l + c, l - c \right) \mod 3,
\quad
l, c \in \{0,1,2\}.
\eg
Matching $l^{(m)}_{w, \bk_*}$ to $l^{(m)}_{u, \bk_*}$ at $K$ and $K'$ fixes $l$ and $c$.
We then have $l^{(3)}_{u, K} + l^{(3)}_{u, K'} = 2l = -l = -l^{(3)}_{w, \Gamma}$ mod 3.
Thus,
\bg
l^{(3)}_{u, \Gamma} - l^{(3)}_{w, \Gamma}
= l^{(3)}_{u, \Gamma} + l^{(3)}_{u, K} + l^{(3)}_{u, K'} 
= \mc C_u \mod 3.
\eg
Likewise, the mismatch can instead be concentrated at $K$ or $K'$.

\paragraph{$C_4$ symmetry:}
The concentration depends on the symmetry orbit on which the mismatch is placed.
$\Gamma$ and $M$ each has a little group $C_4$, and the full value of $\mc C_u \pmod4$ can be concentrated at either point.
On the other hand, each HSM in $\{X,Y\}$ has a little group $C_2$.
Since $\nu (X) = \nu (Y)$ by $C_4$, their total contribution modulo 4 is $\nu (X) + \nu (Y) \bmod 4 \in \{0,2\}$.
Consequently, the mismatch can be concentrated entirely on $\{X,Y\}$ only when $\mc C_u \bmod 4 \in \{0, 2\}$.

The allowed reference angular momenta can be written as
\bg
\left( l^{(4)}_{w, \Gamma}, l^{(4)}_{w, M}, l^{(2)}_{w, X} \right)
= \left( l, l+2a, l+a \right),
\quad
l \in \{0,1,2,3\},
\quad
a \in \{0,1\}.
\eg
To concentrate the mismatch at $\Gamma$, we match angular momenta of $\ket{u (\bk)}$ and $\ket{w (\bk)}$ at $M$ and $X$ such that $l+2a = l^{(4)}_{u,M}$ and $l+a = l^{(2)}_{u,X}$.
Thus, $l = 2 l^{(2)}_{u, X} - l^{(4)}_{u, M} = -2 l^{(2)}_{u, X} - l^{(4)}_{u, M} \bmod 4$, and the remaining mismatch at $\Gamma$ is
\bg
l^{(4)}_{u, \Gamma} - l^{(4)}_{w, \Gamma}
= l^{(4)}_{u, \Gamma} - l
= l^{(4)}_{u, \Gamma} + l^{(4)}_{u, M} + 2 l^{(2)}_{u, X}
= \mc C_u \mod4.
\eg
Similarly, one can concentrate the mismatch at $M$.

Finally, to concentrate the mismatch on $\{X,Y\}$, we match the angular momenta at $\Gamma$ and $M$.
Thus, $l = l^{(4)}_{u, \Gamma}$ and $l+2a = l^{(4)}_{u, M}$.
This is possible only when $l^{(4)}_{u, M} \pm l^{(4)}_{u,\Gamma}$ is even, which is equivalent to $\mc C_u \bmod 4 \in \{0,2\}$.
In this case, the mod-$4$ total vorticity is
\bg
2 \left( l^{(2)}_{u, X} - l^{(2)}_{w, X} \right)
= 2 l^{(2)}_{u, X} - 2 (l+a)
= 2 l^{(2)}_{u,X} - (l^{(4)}_{u, M} + l^{(4)}_{u, \Gamma})
= 2 l^{(2)}_{u,X} + l^{(4)}_{u, M} + l^{(4)}_{u, \Gamma}
= \mc C_u \mod 4.
\eg

\paragraph{$C_6$ symmetry:}
For $C_6$, angular momenta of reference states can be written as
\bg
\left( l^{(6)}_{w,\Gamma}, l^{(3)}_{w,K}, l^{(2)}_{w,M} \right)
= \left( l, l, l \right)
\quad
l \in \{0,1,\dots,5\}.
\eg
To concentrate the mismatch at $\Gamma$, we choose $l$ such that $l = l^{(3)}_{u, K} \bmod 3$ and $l = l^{(2)}_{u, M} \bmod 2$.
These two conditions always determine a unique $l$ modulo $6$.
Also, $2 l^{(3)}_{u, K} = 2l \bmod 6$ and $3 l^{(2)}_{u, M} = 3l \bmod 6$.
The remaining mismatch is at $\Gamma$ and
\bg
l^{(6)}_{u, \Gamma} - l^{(6)}_{w, \Gamma}
= l^{(6)}_{u, \Gamma} - l
= l^{(6)}_{u, \Gamma} + 5 l
= l^{(6)}_{u, \Gamma} + 2 l^{(3)}_{u, K} + 3 l^{(2)}_{u, M}
= \mc C_u \mod 6.
\eg

To concentrate the mismatch on $\{K,K'\}$, we instead match the reference at $\Gamma$ and $M$.
This requires $l = l^{(6)}_{u, \Gamma}$, $l= l^{(2)}_{u, M}$ mod 2.
Under this condition, $\mc C_u$ is even since $\mc C_u \bmod 2 = l^{(6)}_{u, \Gamma} + 3 l^{(2)}_{u, M} \bmod 2= 4l \bmod 2 = 0$.
Then, the mod-$6$ total vorticity is
\bg
2 \left( l^{(3)}_{u, K} - l^{(3)}_{w, K} \right)
= 2 l^{(3)}_{u, K} - 2 l = 2 l^{(3)}_{u, K} + 4 l
= l^{(6)}_{u, \Gamma} + 2 l^{(3)}_{u, K} + 3 l^{(2)}_{u, M}
= \mc C_u \mod 6.
\eg

Finally, to concentrate the mismatch on the three $M$ points, we match the reference at $\Gamma$ and $K$.
This requires $l = l^{(6)}_{u, \Gamma}$ and $l = l^{(3)}_{u, K} \bmod 3$.
From this $C_u$ is a multiple of $3$ since $\mc C_u \bmod 3 = l^{(6)}_{u, \Gamma} + 2 l^{(3)}_{u, K} = 3l \bmod 3 = 0 \bmod 3$.
The remaining mismatch at $\{M,M',M''\}$ satisfies
\bg
3 \left( l^{(2)}_{u, M} - l^{(2)}_{w, M} \right)
= 3 l^{(2)}_{u, M} - 3l
= 3 l^{(2)}_{u, M} + 3l
= l^{(6)}_{u, \Gamma} + 2 l^{(3)}_{u, K} + 3 l^{(2)}_{u, M}
= \mc C_u \mod 6.
\eg

\section{Model calculations}

\subsection{Qi-Wu-Zhang model}
\label{appsub:qwz}

\begin{figure}[t!]
\centering
\includegraphics[width=0.98\textwidth]{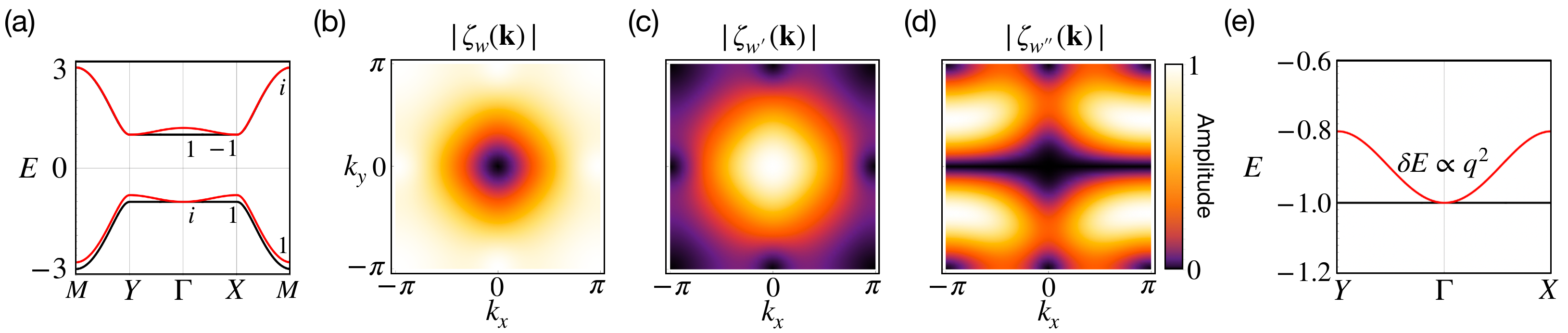}
\caption{Qi-Wu-Zhang model.
(a) Band structure and rotation eigenvalues at the high-symmetry momenta, $\Gamma=(0,0)$ and $M=(\pi,\pi)$ with $C_4$ rotation, and $X=(\pi,0)$ and $Y=(0,\pi)$ with $C_2$ rotation.
The lower band has Chern number $\mc C = 1$.
The black curves show the unperturbed bands, while the red curves show the bands after adding the perturbation $\eta \, \ket{w (\bk)} \bra{w (\bk)}$ with $\ket{w (\bk)} = (1,0)^T$ and $\eta = 0.2$.
(b)-(d) Overlap amplitudes rescaled such that the maximum in each panel is unity.
(b) $|\zeta_w (\bk)|$ for the reference state $\ket{w (\bk)} = (1,0)^T$.
The only zero occurs at $\Gamma$ and has vorticity $\nu (\Gamma) = -1$.
(c) $|\zeta_{w'} (\bk)|$ for $\ket{w' (\bk)} = (0,1)^T$.
The overlap has zeros at $M$, $X$, and $Y$, with vorticities $\nu (M) = 1$ and $\nu(X) = \nu(Y) = -1$.
(d) $|\zeta_{w''} (\bk)|$ for $\ket{w'' (\bk)} = (\sin k_x + i \sin k_y, 1 - \cos k_x)^T$.
The overlap has a line zero $L$ along $k_y=0$ and isolated zeros at $Y$ and $M$, with $\nu (L) = -1$, $\nu (Y) = 1$, and $\nu (M) = -1$.
(e) Enlarged view of the lower band near $\Gamma$ in (a), using the same color convention.
The perturbation produces a quadratic dispersion shift, $\delta E_1 (\bq) \propto q^2$, associated with the $\nu (\Gamma) = -1$ overlap zero shown in (b).}
\label{sfig:qwz}
\end{figure}

\tocless{\subsubsection*}{Symmetry-enforced zeros}
We first consider the two-band Qi-Wu-Zhang model~\cite{qi2006topological} with both orbitals located at $\bx_1 = \bx_2=(0,0)$,
\bg
H (\bk) = \sin k_x \, \sg_x + \sin k_y \, \sg_y + \left( m + \cos k_x + \cos k_y \right) \sg_z.
\eg
We set $m=-1$, for which the lower and upper bands have Chern numbers $\mc C_{u_1} = +1$ and $\mc C_{u_2} = -1$, respectively.
See Fig.~\ref{sfig:qwz}(a) for the band structures and symmetry eigenvalues at the HSMs, $\Gamma = (0,0)$, $X=(\pi,0)$, $Y=(0,\pi)$, and $M=(\pi,\pi)$.
The fourfold rotation is represented by $U_{C_4} (\bk) = {\rm Diag} (1,i)$.
For the lower band $\ket{u_1 (\bk)}$, the rotation eigenvalues correspond to
\bg
l^{(4)}_{u_1, \Gamma} = 1,
\quad
l^{(4)}_{u_1, M} = 0,
\quad
l^{(2)}_{u_1, X} = l^{(2)}_{u_1, Y} = 0,
\eg
according to the definition in Eq.~\eqref{seq:ang_l}.
They satisfy the $C_4$-indicated Chern-number relation, $\mc C_{u_1} = l^{(4)}_{u_1,\Gamma} + l^{(4)}_{u_1,M} + 2l^{(2)}_{u_1,X} = 1 \bmod 4$.
For comparison, the upper band $\ket{u_2 (\bk)}$ has $l^{(4)}_{u_2, \Gamma} = 0$, $l^{(4)}_{u_2, M} = 1$, $l^{(2)}_{u_2, X} = l^{(2)}_{u_2,Y} = 1$, which gives $\mc C_{u_2} =-1$ modulo 4.
In the following, we focus on the lower band $\ket{u_1 (\bk)}$.

We first choose the momentum-independent reference state $\ket{w (\bk)}=(1,0)^T$.
It corresponds to an orbital centered at the unit-cell center with $C_4$ eigenvalue $1$, and hence
\bg
l^{(4)}_{w, \Gamma} = l^{(4)}_{w, M} = l^{(2)}_{w, X} = l^{(2)}_{w,Y} = 0.
\eg
For the lower band, the only symmetry-enforced overlap zero occurs at $\Gamma$, where Eq.~\eqref{seq:vort_rot_w} gives
\bg
\nu (\Gamma) = l^{(4)}_{w, \Gamma} - l^{(4)}_{u_1, \Gamma} = -1 \mod 4.
\eg
Evaluating the local vorticity using the discretized formula [Eq.~\eqref{seq:vort_w_numeric}], we find $\nu (\Gamma)=-1$.
There are no other overlap zeros [Fig.~\ref{sfig:qwz}(b)], and therefore $\nu_{\rm tot} [\zeta_w] = -1 = -\mc C_{u_1}$ in agreement with the vorticity sum rule.
A closely related analysis of the QWZ model was presented in Ref.~\cite{mera2022singular}, where zeros of a projected reference state were studied using singular connections.

We next choose $\ket{w' (\bk)} = (0,1)^T$.
This orbital is also centered at the unit-cell center but has $C_4$ eigenvalue $i$.
Thus,
\bg
l^{(4)}_{w', \Gamma} = l^{(4)}_{w', M} = 1,
\quad
l^{(2)}_{w', X} = l^{(2)}_{w', Y} = 1.
\eg
For the lower band, rotation symmetry therefore requires zeros at $M$, $X$, and $Y$ [Fig.~\ref{sfig:qwz}(c)].
Their vorticities are constrained by
\bg
\nu(M)=1 \mod 4, \quad
\nu(X) = \nu(Y) = 1 \mod 2.
\eg
The discretized calculation yields the corresponding integer vorticities, $\nu (M) = 1$, $\nu (X) = \nu (Y) = -1$.
Consequently, $\nu_{\rm tot} [\zeta_{w'}] = \nu(M) + \nu(X) + \nu(Y) = 1-1-1 = -1 = -\mc C_{u_1}$ in agreement with the vorticity sum rule.

\tocless{\subsubsection*}{Noncontractible line zero}
We illustrate the treatment of a noncontractible line zero discussed in Sec.~\ref{appsub:linezero}.
For this purpose, we choose the following reference state without imposing a rotation constraint:
\bg
\ket{w'' (\bk)} = \left( \sin k_x + i \sin k_y, 1 - \cos k_x \right)^T.
\eg
The corresponding overlap with the lower band has a noncontractible line zero $L$ along $k_y=0$, together with isolated point zeros at $Y$ and $M$ [Fig.~\ref{sfig:qwz}(d)].
The isolated zeros can be treated in the usual way, giving $\nu (Y)=1$ and $\nu (M)=-1$.
Their vorticities modulo $2$ can also be understood from $C_2$ symmetry.
For the line zero, we apply the construction of Sec.~\ref{appsub:linezero} and obtain $\nu (L) = -1$.
Including both the isolated and extended zero sets, the total vorticity is therefore
\bg
\nu_{\rm tot} [\zeta_{w''}] = \nu(L) + \nu(Y) + \nu(M) = -1+1-1 = -1 = -\mc C_{u_1},
\eg
again satisfying the vorticity sum rule.
\\

\tocless{\subsubsection*}{Dispersion shift near the overlap zero}
We finally illustrate how the local structure of an overlap zero appears in the dispersion shift by a perturbation constructed from the reference state.
Returning to $\ket{w (\bk)} = (1,0)^T$, we add a perturbation,
\bg
H_\eta (\bk) = H (\bk) + \eta \, \ket{w (\bk)} \bra{w (\bk)}
= H (\bk) + \eta \bpm 1 & 0 \\ 0 & 0 \epm.
\eg
Here, we set $\eta>0$.
The resulting change in the band structure is shown in Fig.~\ref{sfig:qwz}(e).
Since the overlap $\zeta_w (\bk)$ vanishes at $\Gamma$, the energy shift of the lower band also vanishes there.
To determine its leading behavior, we expand around $\Gamma$ by writing $\bk = \bq$:
\bg
H_\eta (\bq) = \frac{\eta}{2} \sg_0 + q_x \sg_x + q_y \sg_y + \left( 1 + \frac{\eta}{2} - \frac{q^2}{2} \right) \sg_z + O(q^3),
\eg
where $q^2=q_x^2+q_y^2$.
The lower-band energy is
\bg
E_1 (\bq; \eta) = -1 + \frac{\eta}{2 (2 + \eta)} q^2 + O(q^4).
\eg
Comparing with unperturbed lower-band energy, $E_1 (\bq; \eta=0) = -1 + O(q^4)$, we have
\bg
\delta E_1 (\bq) := E_1 (\bq; \eta) - E_1 (\bq; 0)
= \frac{\eta}{2 (2 + \eta)} q^2 + O(q^4).
\eg
For a weak perturbation, $\delta E_1 (\bq) = \frac{\eta}{4} q^2 + O(\eta^2 q^2, q^4)$.
This agrees with the first-order perturbative theory,
\bg
\delta E_1 (\bk) = \eta |\zeta_w(\bk)|^2 + O(\eta^2).
\eg
We now verify directly that $|\zeta_w (\bq)|^2 = \frac{1}{4} q^2 + O(q^4)$.
We write $\bb d (\bk) = \left( \sin k_x, \sin k_y, -1 + \cos k_x + \cos k_y \right)$ and $\ep (\bk) = |\bb d(\bk)|$.
A gauge choice for the lower band that is smooth around $\Gamma$ is
\bg
\ket{u_1 (\bk)} = \frac{1}{\sqrt{2 \ep (\bk) ( \ep (\bk) + d_3 (\bk) )}}
\left( - d_1 (\bk) + i d_2 (\bk), \ep (\bk) + d_3 (\bk) \right)^T.
\eg
Near $\Gamma$, setting $\bk=\bq$, we have
\bg
d_1 (\bq) = q_x + O(q^3), \quad
d_2 (\bq) = q_y + O(q^3), \quad
d_3 (\bq) = 1 - \frac{q^2}{2} + O(q^4),
\quad
\ep (\bq) = 1 + O (q^4).
\eg
Hence, $\ket{u_1 (\bq)} = \frac{1}{2} \left(-q_x + i q_y + O(q^3), 2 + O(q^2) \right)^T$.
For $\ket{w (\bk)} = (1,0)^T$, this gives $\zeta_w (\bq) = -\frac{1}{2} (q_x - i q_y) + O(q^3)$.
Therefore, $|\zeta_w (\bq)|^2 = \frac{1}{4} q^2 + O(q^4)$ in agreement with the dispersion shift above.
The leading term $\zeta_w (\bq) \propto q_x - i q_y$ also directly confirms $\nu (\Gamma) = -1$.
\\

\subsection{Checkerboard-lattice model}

\begin{figure}[t!]
\centering
\includegraphics[width=0.98\textwidth]{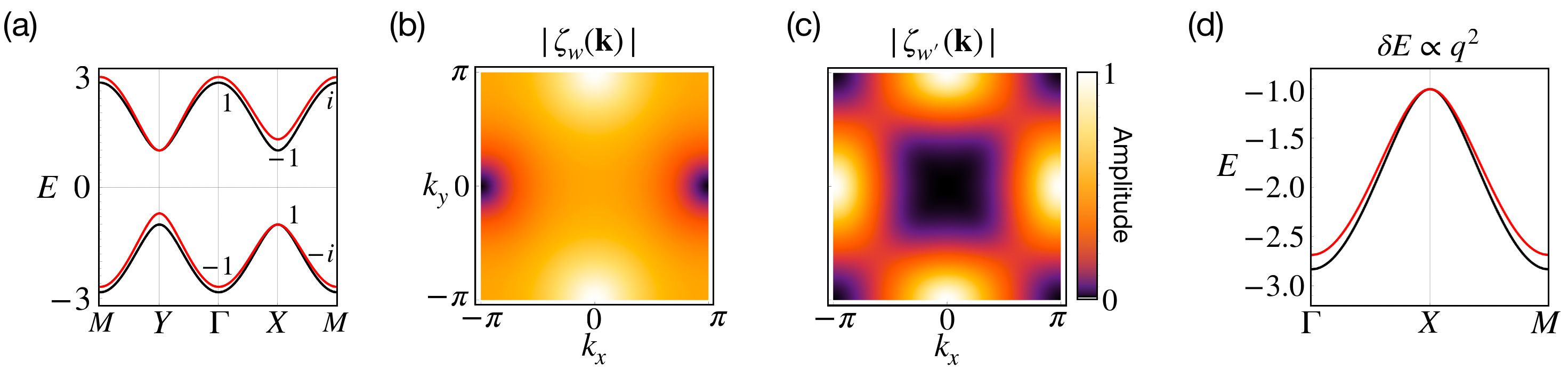}
\caption{Checkerboard-lattice model.
(a) Band structure and rotation eigenvalues at the high-symmetry momenta, $\Gamma=(0,0)$ and $M=(\pi,\pi)$ with $C_4$ rotation, and $X=(\pi,0)$ and $Y=(0,\pi)$ with $C_2$ rotation.
The lower band has Chern number $\mc C = 1$.
The black curves show the unperturbed bands, while the red curves show the bands after adding the perturbation $\eta \, \ket{w (\bk)} \bra{w (\bk)}$ with $\ket{w (\bk)} = (e^{-ik_x/2},0)^T$ and $\eta = 0.3$.
(b)-(c) Overlap amplitudes rescaled such that the maximum in each panel is unity.
(b) $|\zeta_w (\bk)|$ for the reference state $\ket{w (\bk)} = (e^{-i k_x/2},0)^T$.
The only zero occurs at $X$ and has vorticity $\nu (X) = -1$.
(c) $|\zeta_{w'} (\bk)|$ for $\ket{w' (\bk)} = (\cos \frac{k_x}{2}, \cos \frac{k_y}{2})^T$.
The overlap has zeros at $\Gamma$ and $M$, with vorticities $\nu (\Gamma) = -2$ and $\nu(M) = 1$.
(d) Enlarged view of the lower band near $X$ in (a), using the same color convention.
The perturbation produces a quadratic dispersion shift, $\delta E_1 (\bq) \propto q^2$, associated with the $\nu (X) = -1$ overlap zero shown in (b).}
\label{sfig:checkerboard}
\end{figure}

\tocless{\subsubsection*}{Model and symmetry}
We next consider the two-band checkerboard-lattice model with basis orbitals at $\bx_1 = (1/2,0)$ and $\bx_2 = (0,1/2)$.
The fourfold rotation $C_4$ acts on these orbitals as $\hat {C_4} \ket{\bR, 1} = \ket{C_4 \bR, 2}$ and $\hat{C_4} \ket{\bR, 2} = \ket{C_4 \bR - \bb a_1, 1}$.
Thus $C_4$ is represented by $U_{C_4} (\bk) = \sg_x$ in momentum space.
The Hamiltonian is
\bg
H (\bk) = \bpm t_2 \cos k_x + t_3 \cos k_y & 2 t_1 e^{i \Phi} \cos \frac{k_x-k_y}{2} + 2 t_1 e^{-i \Phi} \cos \frac{k_x+k_y}{2} \\
2 t_1 e^{-i \Phi} \cos \frac{k_x-k_y}{2} + 2 t_1 e^{i \Phi} \cos \frac{k_x+k_y}{2} & t_3 \cos k_x + t_2 \cos k_y \epm,
\eg
where we choose $(t_1, t_2, t_3, \Phi) = (1, -0.5, 0.5, \pi/4)$.
The band structure and symmetry eigenvalues at the HSMs are shown in Fig.~\ref{sfig:checkerboard}(a).
For these parameters, the lower band $\ket{u_1 (\bk)}$ has $\mc C_{u_1} = +1$.
Its angular momenta are $l^{(4)}_{u_1, \Gamma} = 2$, $l^{(4)}_{u_1, M} = 3$, and $l^{(2)}_{u_1, X} = l^{(2)}_{u_1, Y} = 0$.
This satisfies $\mc C = l^{(4)}_{u_1, \Gamma} + l^{(4)}_{u_1, M} + 2 l^{(2)}_{u_1, X} = 1 \bmod 4$.
For comparison, the upper band has $( l^{(4)}_{u_2, \Gamma}, l^{(4)}_{u_2, M}, l^{(2)}_{u_2, X} ) = (0,1,1)$, consistently giving $\mc C_{u_2} = -1 \bmod 4$.
In the following, we focus on the lower band.

\tocless{\subsubsection*}{Constraint on $C_4$-symmetric reference states}
In the language of topological quantum chemistry~\cite{bradlyn2017topological}, the full two-band set forms the elementary band representation (EBR) induced from the $(A)_{2c}$ site-symmetry irreducible representation.
An EBR is an elementary building block of topologically trivial insulators.
When an EBR splits into energetically disconnected components, at least one of the split components must be topological~\cite{bradlyn2017topological,cano2018topology,bradlyn2019disconnected,hwang2026srsi}.
In the present two-band case, each component consists of a single nondegenerate band.
Ref.~\cite{alexandradinata2018no} showed that, for any magnetic space group, an isolated nondegenerate band is either topologically trivial or a Chern band.
Hence, at least one of the two split bands must carry a nonzero Chern number.
Since the total Chern number of the full EBR vanishes, the other band must carry the opposite Chern number.
Therefore, whenever the two bands are separated by a gap, both are Chern bands with opposite Chern numbers.
For the parameters chosen here, the lower and upper bands have $\mc C_{u_1} = +1$ and $\mc C_{u_2} = -1$, respectively.

The same argument applies when we try to choose a normalized reference state $\ket{v (\bk)}$ within this orbital basis while preserving $C_4$ symmetry.
Any such state with a smooth projector corresponds to an isolated symmetry-preserving split part of this two-band EBR and must carry a nonzero Chern number.
Hence, a $C_4$-symmetric normalized reference state with a smooth projector cannot be globally smooth.
If we require a globally smooth reference state, we may therefore either break $C_4$ symmetry and use a normalized state with zero Chern number, or preserve $C_4$ symmetry and use an unnormalized state $\ket{w (\bk)}$, which necessarily has zeros in the present case.

\tocless{\subsubsection*}{Reference state without $C_4$ symmetry}
We first choose $\ket{w (\bk)} = (e^{-i k_x/2}, 0)^T$.
This state can be viewed either as an unnormalized reference state obtained from a single-orbital CLS or as a normalized reference state with zero Chern
number.
It does not preserve the $C_4$ symmetry, but satisfies $U_{C_2} (\bk) \ket{w (\bk)} = e^{-i k_x} \ket{w (-\bk)}$.
The corresponding $C_2$ angular momenta at $\Gamma, X, Y,$ and $M$ are $( l^{(2)}_{w, \Gamma}, l^{(2)}_{w, X}, l^{(2)}_{w, Y}, l^{(2)}_{w, M} ) = (0, 1, 0, 1)$ [see Eq.~\eqref{seq:w_sewing_rot}].
For the lower band, $( l^{(2)}_{u_1, \Gamma}, l^{(2)}_{u_1, X}, l^{(2)}_{u_1, Y}, l^{(2)}_{u_1, M} ) = (0,0,0,1)$.
Hence, the only symmetry-enforced overlap zero occurs at $X$, with $\nu (X) = 1 \bmod 2$.
As shown in Fig.~\ref{sfig:checkerboard}(b), the overlap has a single zero at $X$.
The discretized formula gives the integer vorticity $\nu (X) = -1$, and therefore the sum rule $\nu (X) = -1 = -\mc C_{u_1}$ is satisfied.
\\

\tocless{\subsubsection*}{$C_4$-symmetric reference state with a zero}
We next preserve the $C_4$ symmetry and choose $\ket{w' (\bk)} =(\cos \frac{k_x}{2}, \cos \frac{k_y}{2})^T$.
This reference state necessarily vanishes at $M$, $\ket{w' (M)} = (0,0)^T$, while satisfying $U_{C_4} (\bk) \ket{w' (\bk)} = \ket{w' (C_4 \bk)}$.
The sewing phase is trivial, and thus $( l^{(4)}_{w', \Gamma}, l^{(4)}_{w', M}, l^{(2)}_{w', X} ) = (0,0,0)$.
Note that $l^{(4)}_{w', M}$ is defined from the sewing phase even though $\ket{w' (M)} = 0$, as discussed in Sec.~\ref{appsub:rot_condition_w}.
Comparing the angular momenta with those of the lower band, overlap zeros are enforced at $\Gamma$ and $M$ [Fig.~\ref{sfig:checkerboard}(c)], with the constraints $\nu (\Gamma) = 2 \bmod 4$ and $\nu (M) = 1 \bmod 4$.
The numerical calculation gives $\nu (\Gamma) = -2$ and $\nu (M) = 1$, and thus the sum rule $\nu (\Gamma) + \nu (M) = -1 = - \mc C_{u_1}$ is satisfied.
\\

\tocless{\subsubsection*}{Dispersion shift near the overlap zero}
We finally consider the dispersion shift induced by the perturbation $\eta \, \ket{w (\bk)} \bra{w (\bk)}$ for
$\ket{w (\bk)} = (e^{-i k_x/2}, 0)^T$.
The perturbed Hamiltonian is $H_\eta (\bk) = H (\bk) + \eta \, {\rm Diag} (1,0)$.
The resulting change in the band dispersion is shown in Fig.~\ref{sfig:checkerboard}(d).
To obtain the leading behavior around the overlap zero at $X=(\pi,0)$, we write $\bk=X+\bq$ and expand in $\bq$ while keeping $\eta>0$ fixed.
Up to quadratic order,
\bg
H_\eta (X + \bq) = \bpm 1 + \eta - \frac{q^2}{4} & - \sqrt 2 (q_x - i q_y) \\
- \sqrt{2}(q_x + i q_y) & -1 + \frac{q^2}{4}
\epm +O(q^3).
\label{seq:checker_h_expansion}
\eg
At $X$, the lower-band state is $(0, 1)^T$ with energy $-1$, while the other state has energy $1 + \eta$.
Treating $\bq$ as the perturbative parameter, standard second-order perturbation theory gives
\bg
E_1 (X + \bq; \eta) = -1 + \left[ \frac{1}{4} - \frac{2}{2 + \eta} \right] q^2 + O(q^3).
\eg
Subtracting the unperturbed dispersion, we obtain $\delta E_1 (X + \bq) = E_1 (X + \bq; \eta) - E_1 (X + \bq; 0) = \frac{\eta}{2+\eta} q^2 + O(q^3)$.
For a weak perturbation, $\delta E_1 (X + \bq) = \frac{\eta}{2} q^2 + O(\eta^2 q^2, q^3)$.
This agrees with $\delta E_1 (\bk) = \eta |\zeta_w (\bk)|^2 + O(\eta^2)$.
A gauge choice for the lower band that is smooth around $X$ gives $\zeta_w (X + \bq) = - \frac{1}{\sqrt 2} (q_x - i q_y) + O(q^2)$, and hence $|\zeta_w (X + \bq)|^2 = \frac{1}{2} q^2 + O(q^3)$.
Thus, the overlap zero with $\nu(X)=-1$ produces a quadratic dispersion shift around $X$.
Using $C_2$ symmetry, one can further show that the next corrections in $\delta E_1 (X + \bq)$ and $|\zeta_w (X + \bq)|^2$ are $O(q^4)$.
\\

\subsection{Spin-orbit-coupled Lieb model}

\tocless{\subsubsection*}{Model and symmetry}
The three basis orbitals in the Lieb lattice are located at $\bx_1 = (0,0)$, $\bx_2 = (1/2,0)$, $\bx_3 = (0,1/2)$, which gives $V (\bk) = {\rm Diag} \left(1, e^{i k_x/2}, e^{i k_y/2} \right)$.
The orbital at $\bx_1$ has $C_4$ eigenvalue $+1$, while those at $\bx_2$ and $\bx_3$ have $C_2$ eigenvalue $+1$.
The $C_4$ rotation operator and the Hamiltonian are given by
\bg
U_{C_4} (\bk) = \bpm 1 & 0 & 0 \\ 0 & 0 & 1 \\ 0 & 1 & 0 \epm,
\quad
H (\bk) = \bpm 0 & 2\cos \frac{k_x}{2} & 2\cos \frac{k_y}{2} \\ 2\cos \frac{k_x}{2} & 0 & 4i \lambda \sin \frac{k_x}{2} \sin \frac{k_y}{2} \\ 2\cos \frac{k_y}{2} & -4i \lambda \sin \frac{k_x}{2} \sin \frac{k_y}{2} & 0 \epm,
\eg
with $\lambda>0$, as chosen in the main text.
The Hamiltonian satisfies
\bg
U_{C_4} (\bk) H(\bk) U_{C_4} (\bk)^{-1} = H(C_4 \bk),
\quad
C_4 \bk = (-k_y,k_x).
\eg

\tocless{\subsubsection*}{Linear overlap zero of the lowest Chern band}
We derive the local expansion of the overlap function near the zero at $M=(\pi,\pi)$ discussed in the main text.
We first consider the lowest Chern band $\ket{u_1 (\bk)}$ and choose the reference state $\ket{w_1(\bk)}=(1,0,0)^T$.
According to the discussion in Sec.~\ref{appsub:cls}, this reference state is obtained from the CLS consisting of a single orbital state $\ket{\bR, 1}$.
At the $M$ point, $\zeta_{w_1} (M) = \brk{w_1 (M)}{u_1 (M)} = 0$.
This zero is required by rotation symmetry: the reference state has $l^{(4)}_{w_1, M}=0$, whereas the lowest band has $l^{(4)}_{u_1, M}=3$.
Equation~\eqref{seq:vort_rot_w} gives $\nu (M) = l^{(4)}_{w_1,M} - l^{(4)}_{u_1,M} = 1 \bmod 4$.
We now show explicitly from the local expansion that the vorticity is indeed $+1$.

For a nondegenerate band, standard perturbation theory gives
\bg
\ket{u_n (\bk_* + \bq)} = \ket{u_n (\bk_*)} + \sum_{i=x,y} \, q_i \sum_{m \ne n} \frac{\bra{u_m (\bk_*)} (\der_{k_i} H)(\bk_*) \ket{u_n (\bk_*)}}{E_n (\bk_*) - E_m (\bk_*)} \ket{u_m (\bk_*)} + O (q^2).
\eg
Here, $(\der_{k_i} H) (\bk_*)$ denotes $\der_{k_i} H (\bk)$ evaluated at $\bk = \bk_*$.
A term proportional to $\ket{u_n (\bk_*)}$ can in general appear at first order in $\bq$.
Its real part vanishes by the normalization condition, while the remaining imaginary part can be removed by a smooth $U(1)$ gauge choice.

For a general reference state, $\zeta_w (\bk_* + \bq)-\zeta_w (\bk_*)$ is given by
\bg
\sum_i \, q_i \brk{\der_{k_i} w (\bk_*)}{u_n (\bk_*)}
+ \sum_i \, q_i \sum_{m \ne n} \, \frac{\bra{u_m (\bk_*)} (\der_{k_i} H) (\bk_*) \ket{u_n (\bk_*)}} {E_n (\bk_*) - E_m (\bk_*)} \brk{w (\bk_*)}{u_m (\bk_*)} + O(q^2).
\label{seq:zeta_expansion_pert}
\eg
For the present case, $\ket{w_1 (\bk)}$ is independent of $\bk$ and $\zeta_{w_1} (M)=0$, so Eq.~\eqref{seq:zeta_expansion_pert} reduces to
\bg
\zeta_{w_1} (M + \bq) = \sum_{i=x,y} \, q_i \sum_{m=2,3} \, \frac{\bra{u_m (M)} (\der_{k_i} H)(M) \ket{u_1 (M)}}{E_1 (M) - E_m (M)} \brk{w_1 (M)}{u_m (M)} + O(q^2).
\label{seq:lieb_overlap_1}
\eg

At $M$, the eigenstates and energies can be chosen as
\bg
\ket{u_1 (M)} = \frac{1}{\sqrt 2} (0, -i, 1)^T,
\quad
\ket{u_2 (M)} = (1, 0, 0)^T,
\quad
\ket{u_3 (M)} = \frac{1}{\sqrt 2} (0, i, 1)^T,
\label{seq:lieb_eigvec}
\eg
and $\left( E_1 (M), E_2 (M), E_3 (M) \right) = (-4 \lambda, 0, 4 \lambda)$.
Since $\ket{w_1 (\bk)} = (1,0,0)^T = \ket{u_2 (M)}$, only the $m=2$ term contributes to Eq.~\eqref{seq:lieb_overlap_1}.
Evaluating the derivatives of the Hamiltonian at $M$, we obtain
\bg
\zeta_{w_1} (M + \bq) = - \frac{i}{4 \sqrt{2} \lambda} (q_x + i q_y) + O(q^2) = - \frac{i}{4 \sqrt{2} \lambda} q_+ + O(q^2).
\label{seq:lieb_zeta1}
\eg
Thus, the leading term is proportional to $q_+$, giving $\nu (M) = +1$, in agreement with the rotation-symmetry constraint above.

\tocless{\subsubsection*}{Quadratic overlap zero of the trivial flat band}
We next consider the trivial flat band $\ket{u_2 (\bk)}$ with $E_2 (\bk)=0$. 
For this band, we choose
\bg
\ket{w_2 (\bk)} = V(-\bk) \ket{\cm{w}_2 (\bk)},
\quad
\ket{\cm{w}_2 (\bk)} = (0, 1 + e^{-i k_y}, - 1 - e^{-i k_x} )^T.
\eg
As discussed in Sec.~\ref{appsub:cls}, $\ket{\cm{w}_2 (\bk)}$ is the Fourier transform of the CLS, $\ket{w_2 (\bR)} = \ket{\bR, 2} + \ket{\bR + \bb a_2, 2} - \ket{\bR, 3} - \ket{\bR + \bb a_1, 3}$.
In this case, the flat-band eigenstate is known exactly:
\bg
\ket{u_2 (\bk)} = \mc N (\bk) \left(2 i \lambda \sin \tfrac{k_x}{2} \sin \tfrac{k_y}{2}, \cos \tfrac{k_y}{2}, - \cos \tfrac{k_x}{2} \right)^T,
\label{seq:lieb_u2}
\eg
where $\mc N(\bk)$ is the normalization factor.
Near $M$, the unnormalized vector in Eq.~\eqref{seq:lieb_u2} becomes $\left( 2i \lambda, -\frac{q_y}{2}, \frac{q_x}{2} \right)^T + O(q^2)$.
Fixing the overall phase such that $\mc N (M)$ is real, we have $\mc N (M+\bq) \simeq \frac{1}{2 \lambda} + O (q^2)$.
Using $V(-\bk)$, the reference state can be written as
\bg
\ket{w_2 (\bk)} = V(-\bk) (0, 1 + e^{-i k_y}, - 1 - e^{-i k_x} )^T
= 2 e^{-i \frac{k_x}{2} - i \frac{k_y}{2}} \left( 0, \cos \tfrac{k_y}{2}, - \cos \tfrac{k_x}{2} \right)^T.
\eg
The overlap is therefore
\bg
\zeta_{w_2} (\bk) = \brk{w_2 (\bk)}{u_2 (\bk)}
= 2 \mc N(\bk) e^{i \frac{k_x}{2} + i \frac{k_y}{2}} \left( \cos^2 \tfrac{k_x}{2} + \cos^2 \tfrac{k_y}{2} \right).
\eg
Expanding around $M=(\pi,\pi)$ gives
\bg
\zeta_{w_2} (M + \bq) = -\frac{1}{4\lambda} (q_x^2 + q_y^2) + O(q^3).
\eg
Thus, the overlap vanishes quadratically at $M$, but its leading term is proportional to $q^2=q_+q_-$ and carries zero vorticity.
This illustrates that the order of an overlap zero and its vorticity are distinct quantities.
\\

\tocless{\subsubsection*}{Dispersion shift near the overlap zero}
We finally consider the dispersion shift induced by the perturbation $\eta \, \ket{w_1 (\bk)} \bra{w_1 (\bk)}$ with $\ket{w_1 (\bk)} = (1,0,0)^T$.
The perturbed Hamiltonian is $H_\eta (\bk) = H (\bk) + \eta \, {\rm Diag} (1,0,0)$.
To obtain the leading dispersion around $M$, we write
$\bk = M + \bq$ and expand the Hamiltonian in $\bq$ while keeping $\eta$ fixed:
\bg
H_\eta (M + \bq) = \bpm
\eta & -q_x & -q_y \\
- q_x & 0 & 4i \lambda - \frac{i \lambda}{2} q^2 \\
- q_y & - 4i \lambda + \frac{i \lambda}{2} q^2 & 0 \epm + O(q^3)
= H^{(0)} + H^{(1)} + H^{(2)} + O(q^3).
\eg
Here, $H^{(0)} = H_\eta(M)$, and $H^{(1)}$ and $H^{(2)}$ denote the terms linear and quadratic in $\bq$, respectively.
The eigenstates and energies of $H^{(0)}$ are given by
\bg
\ket{u_1^{(0)}} = \frac{1}{\sqrt 2} (0, -i, 1)^T,
\quad
\ket{u_2^{(0)}} = (1, 0, 0)^T,
\quad
\ket{u_3^{(0)}} = \frac{1}{\sqrt 2}(0, i, 1)^T,
\quad
\left(E_1^{(0)}, E_2^{(0)}, E_3^{(0)} \right) = (-4 \lambda, \eta, 4 \lambda).
\eg
So, we take $0 < \eta < 4 \lambda$, so that the ordering of the three states at $M$ is unchanged.
Treating $\bq$ as the perturbative parameter, the lowest-band energy up to $O(q^2)$ follows from standard nondegenerate perturbation theory,
\bg
E_n (M + \bq; \eta) = E_n^{(0)} + \bra{u_n^{(0)}} ( H^{(1)} + H^{(2)} ) \ket{u_n^{(0)}} + \sum_{m \ne n} \, \frac{ |\bra{u_m^{(0)}} H^{(1)} \ket{u_n^{(0)}}|^2}{E_n^{(0)} - E_m^{(0)}} + O(q^3).
\eg
For the present case, $E_1 (M + \bq; \eta) = -4 \lambda + \left[ \frac{\lambda}{2} - \frac{1}{2 (4 \lambda + \eta)} \right] q^2 + O(q^3)$.
Subtracting the unperturbed dispersion, we obtain $\delta E_1 (M + \bq) = E_1 (M + \bq; \eta) - E_1 (M + \bq; 0) = \frac{\eta}{8 \lambda (4 \lambda + \eta)} q^2 + O(q^3)$.
For a weak perturbation,
\bg
\delta E_1(M + \bq) = \frac{\eta}{32 \lambda^2} q^2 + O(\eta^2 q^2, q^3).
\eg
This agrees with the first-order perturbative relation, $\delta E_1 (\bk) = \eta |\zeta_{w_1} (\bk)|^2 + O(\eta^2)$, and Eq.~\eqref{seq:lieb_zeta1}.
\\

\section{One-dimensional inversion and the Berry phase}
Our construction based on overlap zeros also applies to topological invariants other than the Chern number.
As a simple example, we consider a one-dimensional (1D) inversion-symmetric system.
Let $\ket{u (k)}$ and $\ket{v (k)}$ be normalized
states defining smooth projectors over the 1D BZ, and let inversion $\mc I$ act as $U_{\mc I} (k) \ket{a (k)} = \xi_{a, \mc I} (k) \ket{a (-k)}$, $a=u,v$.
The construction extends analogously to an unnormalized reference state $\ket{w (k)}$.
At the inversion-invariant momenta $k_* \in \{0, \pi \}$, the sewing phases reduce to the inversion eigenvalues, $\xi_{a, \mc I} (k_*) = \{+1, -1\}$.
The overlap $\zeta_v (k) = \brk{v (\bk)}{u (\bk)}$ transforms as $\zeta_v (-k) = \frac{\xi_{v, \mc I} (k)}{\xi_{u, \mc I} (k)} \zeta_v (k)$.
Therefore, a mismatch between the inversion eigenvalues of the two states enforces an overlap zero at the corresponding inversion-invariant momentum.

As a 1D counterpart of the vorticity $\nu$, we consider the order $p_{k_*}$ of the leading nonzero term in the overlap.
Choosing smooth gauges for both states in a neighborhood of $k_*$ and writing $k = k_* + q$, we write the leading expansion as $\zeta_v (k_* + q) \propto q^{p_{k_*}} + O(q^{p_{k_*} + 1})$.
The inversion constraint gives
\bg
(-1)^{p_{k_*}} = \frac{\xi_{v, \mc I}(k_*)}{\xi_{u, \mc I}(k_*)}.
\eg
Thus, an inversion-eigenvalue mismatch requires a zero of odd order, whereas equal inversion eigenvalues require $p_{k_*}$ to be even.
In the latter case, $p_{k_*} = 0$ corresponds to a nonzero overlap, while an accidental zero, if present, has even order.

The Berry phase $\Phi_a = \oint_{\rm BZ} \mc A_a \bmod{2 \pi}$ is quantized by inversion symmetry according to
\bg
e^{i \Phi_a} = \xi_{a, \mc I} (0) \xi_{a,\mc I} (\pi).
\eg
Combining this relation with the local constraints above gives
\bg
e^{i (\Phi_v - \Phi_u)} = (-1)^{p_0 + p_\pi}.
\eg
Zeros away from the inversion-invariant momenta $k_*$ occur in inversion-related pairs and therefore do not change the parity of the total zero order.
Consequently, $\Phi_v - \Phi_u = \pi (p_0 + p_\pi)$ mod $2 \pi$.
Thus, in 1D, the parity of the total order of the overlap zeros diagnoses the difference between the inversion-quantized Berry phases of the target and reference states.
If all zeros are simple, this reduces to counting the number of overlap zeros modulo 2, as in the zero-counting approach of Ref.~\cite{hwang2026exciton}.

\section{Relation to the quantum metric}
We first consider normalized reference states.
At a zero $\bk_*$ of $\zeta_v (\bk) = \brk{v (\bk)}{u (\bk)}$, we have $\brk{v}{u}=0$.
Here and below, quantities without an explicit momentum argument are evaluated at $\bk_*$.
Therefore,
\bg
\der_{k_x} \zeta_v = \brk{\der_{k_x} v}{u} + \brk{v}{\der_{k_x} u}
= \bra{\der_{k_x} v} \left( \mathds{1} - \ket{v} \bra{v} \right) \ket{u} + \bra{v} \left( \mathds{1} - \ket{u} \bra{u} \right) \ket{\der_{k_x} u}
= \bra{\der_{k_x} v} Q_v \ket{u} + \bra{v} Q_u \ket{\der_{k_x} u}.
\eg
Here, we defined $Q_a = \mathds{1} - \ket{a} \bra{a} = \mathds{1} - P_a$ for $a=u,v$, such that $Q_a \ket{a} =0$ and $P_a \ket{a} = \ket{a}$.
Using the triangle inequality $|\der_{k_x} \zeta_v| \le |\bra{\der_{k_x} v} Q_v \ket{u}| + |\bra{v} Q_u \ket{\der_{k_x} u}|$.
Also, $|\bra{\der_{k_x} v} Q_v \ket{u}|^2 = \bra{\der_{k_x} v} Q_v \ket{u} \bra{u} Q_v \ket{\der_{k_x} v} \le |Q_v \ket{\der_{k_x} v}|^2 = \bra{\der_{k_x} v} Q_v \ket{\der_{k_x} v}$, and similarly $|\bra{v} Q_u \ket{\der_{k_x} u}|^2 \le \bra{\der_{k_x} u} Q_u \ket{\der_{k_x} u}$.
The quantum metric is defined as
\bg
g_{a, ij} (\bk) = {\rm Re} \, \bra{\der_{k_i} a (\bk)} \left( \mathds{1} - \ket{a (\bk)} \bra{a (\bk)} \right) \ket{\der_{k_j} a (\bk)}.
\eg
Thus,
\bg
|\der_{k_x} \zeta_v| \le \sqrt{g_{v, xx}} + \sqrt{g_{u, xx}}.
\label{seq:metric_bound_1}
\eg
For a constant normalized reference state, $g_{v,ij} = 0$, and Eq.~\eqref{seq:metric_bound_1} reduces to $|\der_{k_x} \zeta_v (\bk_*)|^2 \le g_{u, xx} (\bk_*)$.
The same argument can be applied along other momentum directions.
For example, defining $\der_\pm = \frac{1}{\sqrt 2}(\der_{k_x} \pm \der_{k_y})$, one obtains
\bg
|\der_\pm \zeta_v (\bk_*)| \le \sqrt{\frac{g_{v, xx} + g_{v, yy} \pm 2 g_{v, xy}}{2}} + \sqrt{\frac{g_{u, xx} + g_{u, yy} \pm 2 g_{u, xy}}{2}}.
\label{seq:metric_bound_2}
\eg
More generally, consider a unit vector $\bb n$ and define $k_n = \bb n \cdot \bk$ and $\der_{k_n} = n_1 \der_{k_x} + n_2 \der_{k_y}$.
Then, $|\der_{k_n} \zeta_v| \le \sqrt{n_i \, g_{v, ij} \, n_j} + \sqrt{n_i \, g_{u, ij} \, n_j}$.
Eq.~\eqref{seq:metric_bound_2} is reproduced for $\bb n = \frac{1}{\sqrt 2} (1, \pm 1)$.
The relation between these bounds and the dispersion shift is discussed below together with unnormalized reference states.

We next consider an unnormalized reference state $\ket{w (\bk)}$.
Suppose first that $\ket{w (\bk_*)} \ne 0$ at an overlap zero.
In a neighborhood of $\bk_*$, we may define the normalized reference state $\ket{v (\bk)} = \ket{w (\bk)} \sqrt{\brk{w (\bk)}{w (\bk)}}$.
Writing $\ket w = \mc N_w \ket v$, where $\mc N_w = \sqrt{\brk{w}{w}} > 0$, we have $\zeta_w = \mc N_w \zeta_v$.
Since $\zeta_v (\bk_*) = 0$, $\der_{k_x} \zeta_w (\bk_*) = \mc N_w (\bk_*) \der_{k_x} \zeta_v (\bk_*)$.
Then the preceding result for the normalized reference states gives $|\der_{k_x} \zeta_w| \le \mc N_w \left( \sqrt{g_{v, xx}} + \sqrt{g_{u, xx}} \right)$.

The situation is different when $\ket{w (\bk_*)}=0$.
In this case, $\der_{k_i} \zeta_w (\bk_*) = \brk{ \der_{k_i} w}{u}$, so the leading derivative of the overlap is determined by the local variation of the reference state and does not in general yield a direct bound on the quantum metric of $\ket u$.
This is also natural from the viewpoint of the reference-state projector $P_w (\bk) = \frac{\ket{w (\bk)}\bra{w (\bk)}}{\brk{w (\bk)}{w (\bk)}}$, which is undefined at $\bk_*$ and need not admit a smooth extension there, even when $\ket{w (\bk)}$ itself is smooth.
Hence, a quantum metric associated with $P_w$ cannot in general be defined at such a zero.

Nevertheless, the local overlap structure is directly reflected in the dispersion shift generated by $H_\eta (\bk) = H (\bk) + \eta \, \ket{w (\bk)} \bra{w (\bk)}$.
To first order in $\eta$, $\delta E_u (\bk) = \eta |\zeta_w (\bk)|^2 + O(\eta^2)$.
At an overlap zero $\bk_*$, this gives
\bg
\frac{1}{2 \eta} \der_{k_n}^2 \delta E_u = |\der_{k_n} \zeta_w|^2 + O(\eta),
\label{seq:dispersion_zeta}
\eg
which can be combined with the quantum-metric bound derived above when $\ket {w (\bk_*)} \ne 0$.
In particular, for a momentum-independent reference state $\ket w$, $g_{v, ij}=0$.
For example, $\frac{1}{2 \eta \mc N_w^2} \der_{k_x}^2 \delta E_u = |\der_{k_x} \zeta_v|^2 + O(\eta) \le g_{u,xx} + O(\eta)$.

For a two-band system, the constant-probe result becomes particularly simple.
At the overlap zero, $\ket u$ and $\ket v$ span the
two-dimensional Hilbert space, so that $Q_u = \ket v \bra v$.
Consequently, $g_{u, ij} (\bk_*) = {\rm Re} \, \left[ (\der_{k_i} \zeta_v)^* (\der_{k_j} \zeta_v) \right]_{\bk_*}$ and Eq.~\eqref{seq:dispersion_zeta} yields
\bg
\frac{1}{2 \eta \mc N_w^2} \der_{k_i} \der_{k_j} \delta E_u (\bk_*) = g_{u, ij} (\bk_*) + O(\eta).
\eg

\section{Connections and overlap vorticity}
We first recall the notion of a line bundle associated with an isolated Bloch band.
Let $\ket{u (\bk)}$ be a normalized eigenstate and $P_u (\bk) = \ket{u (\bk)} \bra{u (\bk)}$ the corresponding projector.
At each momentum $\bk$, the eigenstate defines a one-dimensional complex vector space $L_\bk$, spanned by $\ket{u (\bk)}$: $L_\bk = {\rm Ran} \, P_u (\bk) = \{c \ket{u(\bk)} | c \in \C \}$.
Here, ${\rm Ran} \, P_u (\bk)$ denotes the range of the projector, i.e., the set of vectors $P_u (\bk) \ket{\phi}$ for arbitrary $\ket{\phi}$.
The collection of these spaces over the BZ forms a complex line bundle $L$.
A section of $L$ specifies a vector $\ket{s (\bk)} \in L_\bk$ at each momentum $\bk$, allowing us to describe a state that varies continuously across the BZ.
In a local patch where a smooth eigenstate $\ket{u (\bk)}$ can be chosen, a smooth section takes the form $\ket{s (\bk)} = f (\bk) \ket{u (\bk)}$, where $f (\bk)$ is a smooth complex scalar that specifies the vector $\ket{s (\bk)}$ within $L_\bk$.
Unlike a normalized eigenstate, a section need not have unit norm and may vanish at some momenta.

At different momenta, the vector spaces $L_\bk$ are spanned by the corresponding eigenstates $\ket{u (\bk)}$, which generally vary with $\bk$.
Thus, comparing vectors at neighboring momenta requires an additional prescription.
A connection $\nabla$ provides such a prescription through a covariant derivative, which maps a section to a one-form taking values in $L_\bk$ at each momentum.
It satisfies the Leibniz rule
\bg
\nabla (g \ket s) = dg \, \ket s + g \nabla \ket s
\eg
for any smooth complex function $g$ and section $\ket s$.
A smooth local eigenstate $\ket{u (\bk)}$ is itself a section, and thus the Leibniz rule gives
\bg
\nabla \ket{s (\bk)} = \nabla \left( f (\bk) \ket{u (\bk)} \right)
= df (\bk) \ket {u (\bk)} +f (\bk) \nabla {u (\bk)}.
\eg
Since $\nabla \ket{u (\bk)}$ is a one-form taking values in the one-dimensional space $L_\bk$, it can be written as
\bg
\nabla \ket{u (\bk)} = -i \mc A (\bk) \ket{u (\bk)},
\eg
where $\mc A (\bk) = \sum_i \, \mc A_i (\bk) dk_i$ is the local connection one-form.
Then,
\bg
\nabla \ket{s (\bk)} = \left[ df (\bk) - i f (\bk) \mc A (\bk) \right] \ket{u (\bk)}.
\eg

\subsection{Berry and singular connections}
A familiar example is the Berry connection~\cite{berry1984quantal}, defined by projecting the ordinary derivative onto the band subspace,
\bg
\nabla^B \ket s = P_u \ket {d s}.
\eg
For $\ket{s (\bk)} = f (\bk) \ket {u (\bk)}$, this gives $\nabla^B \ket {s (\bk)} = \left[ df (\bk) + f (\bk) \brk{u (\bk)}{du (\bk)} \right] \ket {u (\bk)}$, where $\mc A^B (\bk) = i \brk{u (\bk)}{du (\bk)}$ is the usual Berry connection one-form.
The projector $P_u (\bk)$ is smooth for an isolated band.
It is also globally defined, i.e., well defined throughout the BZ rather than only within local patches, because it is invariant under $\ket{u(\bk)} \to e^{i \lambda (\bk)} \ket{u (\bk)}$.
Thus, the Berry connection $\nabla^B = P_u d$ is globally defined without requiring a globally smooth eigenstate.
Another connection can be constructed by specifying a smooth global section and requiring it to be covariantly constant wherever it is nonzero.

The singular-connection approach of Ref.~\cite{mera2022singular} uses this construction.
Consider a smooth global section $\ket{s (\bk)}$ of $L$, which may vanish at certain momenta.
A connection $\nabla^S$ is defined away from these zeros by requiring
\bg
\nabla^S \ket{s (\bk)} = 0.
\eg
Writing $\ket{s (\bk)} = f (\bk) \ket{u (\bk)}$ within a local patch and using the Leibniz rule gives $\mc A^S (\bk) = -i d \log f (\bk)$, where $f (\bk) \ne 0$.
Since $\ket{s (\bk)}$ is globally defined, it must represent the same vector under a local gauge transformation $\ket{u (\bk)} \to e^{i \lambda (\bk)} \ket{u (\bk)}$.
Thus, $f (\bk) \to e^{-i \lambda (\bk)} f (\bk)$, which implies $\mc A^S (\bk) \to \mc A^S (\bk) - d \lambda (\bk)$.
This is precisely the gauge transformation law of a connection one-form, ensuring that $\nabla^S$ is well defined across different local patches away from the zeros of $s$.
Without a globally defined $s$, the local expressions $\mc A^S = -i d \log f$ would not necessarily satisfy this transformation law on overlapping patches.
The resulting connection is flat, i.e., its curvature vanishes away from the zeros of $s$.
Its singular curvature is concentrated at these zeros, and its integral over the BZ gives the Chern number~\cite{mera2022singular}.

\subsection{Unnormalized case}
We now discuss our overlap-vorticity formulation in terms of connections.
First, consider the unnormalized-reference case, where $\ket{w (\bk)}$ is a smooth reference state defined globally over the BZ, with the overlap $\zeta_w (\bk) = \brk{w (\bk)}{u (\bk)}$.
We will discuss the normalized-reference case separately below.
Projecting the reference state onto the band gives
\bg
\ket{s_w (\bk)} = P_u (\bk) \ket{w (\bk)} = \zeta_w^* (\bk) \ket{u (\bk)}.
\eg
Since both $P_u (\bk)$ and $\ket{w (\bk)}$ are smooth and globally defined, $\ket{s_w (\bk)}$ is a smooth global section of $L$.
Moreover, its zeros coincide with those of $\zeta_w (\bk)$.
We can therefore define a singular connection $\nabla^S$ away from these zeros.
Applying the singular-connection construction to $\ket{s_w (\bk)}$ gives
\bg
\mc A^S (\bk) = -i d \log \zeta_w^* (\bk) = - d \arg \zeta_w (\bk) - i d \log |\zeta_w (\bk)|,
\eg
away from the overlap zeros.
In our overlap-vorticity formulation, we instead consider the phase winding of $\zeta_w (\bk)$, defined through $d \arg \zeta_w (\bk)$, without introducing a connection [see Eq.~\eqref{seq:vort_v}].
The real part of $\mc A^S (\bk)$ is therefore $-d \arg \zeta_w (\bk)$, while its imaginary part depends on the magnitude of the overlap.
The latter does not contribute to the winding around a closed loop on which $\zeta_w (\bk)$ is nonzero.

Away from the zeros of $\zeta_w$, we can define a connection $\nabla^{\rm UN}$ for the unnormalized case by requiring the normalized section $\ket{\hat s_w (\bk)} = \ket{s_w (\bk)}/ ||s_w (\bk) ||$ to be covariantly constant,
\bg
\nabla^{\rm UN} \ket{\hat s_w (\bk)} = 0.
\eg
Since $\ket{\hat s_w (\bk)} = e^{-i \arg \zeta_w (\bk)} \ket{u (\bk)}$, this gives
\bg
\mc A^{\rm UN} (\bk) = - d \arg \zeta_w (\bk).
\eg
Thus, the phase-winding one-form used in our formulation can also be interpreted, up to a minus sign, as a local connection one-form.

\subsection{Normalized case}
We next consider the normalized-reference case, where $\ket{v (\bk)}$ is an eigenstate of another isolated Bloch band, with projector $P_v (\bk)$ and Chern number $\mc C_v$.
Unlike the unnormalized reference state $\ket{w (\bk)}$, $\ket{v (\bk)}$ need not be globally smooth.
Consequently, $P_u (\bk) \ket{v (\bk)}$ does not generally define a global section of $L$.
Instead, consider the operator
\bg
\mc T (\bk) = P_u(\bk) P_v (\bk) =\ \zeta_v^* (\bk) \ket{u (\bk)} \bra{v (\bk)},
\eg
where $\zeta_v (\bk) = \brk{v (\bk)}{u (\bk)}$.
Since both projectors are globally defined, $\mc T (\bk)$ is also smooth and globally defined.

At each momentum $\bk$, consider linear operators mapping the one-dimensional vector space $L^v_\bk = {\rm Ran} \, P_v (\bk)$ to $L^u_\bk = {\rm Ran} \, P_u (\bk)$.
Since both spaces are one-dimensional, any such operator is of the form $c \ket{u (\bk)}\bra{v (\bk)}$ with $c \in \C$.
These operators therefore form a one-dimensional complex vector space, to which $\mc T(\bk)$ belongs.
As $\bk$ varies, these vector spaces form another line bundle, of which $\mc T (\bk)$ is a smooth global section.
As in the unnormalized-reference case, requiring $\nabla^S \mc T = 0$ gives a singular connection with local one-form $\mc A^S = -i d \log \zeta_v^*$.
This includes a contribution from the magnitude of $\zeta_v$, whereas our overlap-vorticity formulation uses only its phase winding $d \arg \zeta_v (\bk)$.
To obtain a connection one-form corresponding directly to the latter, we can instead define a connection $\nabla^{\rm N}$ by requiring
\bg \nabla^{\rm N} \left( \mc T (\bk) / || \mc T (\bk) || \right) = 0,
\eg
where $||\mc T (\bk)|| = |\zeta_v (\bk)|$.
Since $\mc T (\bk) / ||\mc T (\bk)|| = e^{-i \arg \zeta_v (\bk)} \ket{u (\bk)} \bra{v (\bk)}$, this gives $\mc A^{\rm N} (\bk) = -d \arg \zeta_v (\bk)$.
Unlike $L$, this line bundle has Chern number $\mc C_u - \mc C_v$, consistent with the vorticity sum rule for the normalized-reference case.

In summary, our overlap-vorticity formulation can be interpreted in terms of connections, but its derivation does not rely on introducing them.
This connection-based interpretation is natural because the Chern number is a topological invariant of a line bundle, which can be computed from the curvature of any smooth connection, or from suitably defined singular connections.

\end{document}